\documentclass[useAMS,usenatbib]{mn2e}

\usepackage{times}
\usepackage{graphicx}
\usepackage{ulem}
\usepackage{epsfig}
\usepackage{amssymb}

\usepackage{psfrag}
\def\apj{ApJ}
\def\apjl{ApJ}
\def\mnras{MNRAS}
\def\prd{Phys.~Rev.~D}
\def\nat{Nature}
\def\physrep{Phys.~Rep.}

\def\be{\begin{equation}}
\def\ee{\end{equation}}
\def\bary{\begin{eqnarray}}
\def\eary{\end{eqnarray}}

\def\bi{\begin{itemize}}
\def\ei{\end{itemize}}
\def\lsim{\mathrel{\rlap{\lower3pt\hbox{\hskip1pt$\sim$}}
     \raise1pt\hbox{$<$}}} 
\def\gsim{\mathrel{\rlap{\lower3pt\hbox{\hskip1pt$\sim$}}
     \raise1pt\hbox{$>$}}} 

\def\cal{\it}

\title[GeV - PeV Neutrino Production and Oscillation  in  hidden jets from GRBs]
{GeV - PeV Neutrino Production and Oscillation  in  hidden jets from GRBs}
\author[N. Fraija]%
{Nissim Fraija \thanks{E-mail:nifraija@astro.unam.mx. Luc Binette-Fundaci\'on UNAM Fellow.}\\
Instituto de Astronom\' ia, Universidad Nacional Aut\'onoma de M\'exico, Circuito Exterior, \\C.U., A. Postal 70-264, 04510 M\'exico D.F., M\'exico}

\begin{document}

\maketitle
	
\begin{abstract}
Long gamma-ray bursts have been widely associated with collapsing massive stars in the framework of   collapsar model.    High-energy neutrinos and photons can be produced in the internal shocks of middle relativistic jets from core-collapse supernova.  Although photons can hardly escape, high-energy neutrinos could be the only signature when the jets are hidden.  We show that using suitable parameters, high-energy neutrinos in GeV - PeV range  can be produced in the hidden jet inside the collapsar, thus demonstrating that these objects are candidates to produce neutrinos with energies between  1 - 10 PeV  which were observed with IceCube.  On the other hand, due to matter effects, high-energy neutrinos may oscillate resonantly from one flavor to another before leaving the star. Using two (solar, atmospheric and accelerator parameters) and three neutrino mixing, we study the possibility of resonant oscillation for these neutrinos created in internal shocks. Also we compute the probabilities of neutrino oscillations in the matter at different distances along the jet (before leaving the star) and after in vacuum, on their path to Earth.   Finally, neutrino flavor ratios on Earth are estimated. 
  
\end{abstract}


\section{Introduction}\label{sec-Intro}
A number of different high-energy, $\geq$ 1GeV, neutrino sources have been proposed in literature, that include active galactic nuclei (AGNs)\citep{ste91,sza94,nell93, ato01,alv04}, gamma-ray bursts (GRBs)\citep{wax97,der03,raz04b,mur06,gup07}, supernova remnants \citep{alv02,cos05} and core collapse supernovae \citep{wax01, wan07}, although   long duration GRBs have been found to be tightly connected with core-collapse supernovae \citep{hjo03,sta03}.  Properties of neutrino fluxes, energy range, shape of the energy spectra and flavor content depend on physical conditions in the sources.  
Neutrinos are useful for studying sources, especially when photons cannot escape directly. They could be the only prompt signatures of the "hidden" sources.  These have been associated to core collapse of massive stars leading to supernovae (SNe) of type Ib,c and II with mildly relativistic jets emitted by a central engine, a black hole or a highly magnetized neutron star.  Depending on the initial density and metallicity, the pre-supernova star could have different radii. Type Ic supernovae are believed to be He stars with radius $R_\star\approx $ 10$^{11}$ cm and Supernovae of type II and Ib are thought to have a radius of $R_\star\approx$ 3$\times$10$^{12}$ cm. \\
Recently, IceCube reported the detection of two neutrino-induced events with energies between 1- 10 PeV \citep{aar13}.  These events have been discussed as having an extragalactic origin, for instance; GRBs\citep{cho12}  and low-luminosity GRBs \citep{liu13}. On the other hand, high-energy neutrinos are produced in the decay of charged pions and muons when energetic protons in the jet interact with synchrotron thermalized photons or nucleons/mesons (pp, pn)/($\pi$, K) in the shocks.  For internal shocks, synchrotron radiation and the number density of particles could be calculated with enough accuracy if we know the distribution of the magnetic field and the particle momentum in the shocked region.  These quantities are calculated using the energy equipartition hypothesis through the equipartition parameters; electron equipartition ($\epsilon_e=U_e/U$) and magnetic equipartition $\epsilon_B=U_B/U$\citep{mes98}. Many authors \citep{barn12,fra12,sac12,kum10,she10} have estimated these parameters to be  $\epsilon_e\simeq$ 0.1, and  0.1$\leq \epsilon_B\leq 10^{-4}$, to obtain a good description of more than a dozen of GRBs.\\
On the other hand, the neutrino flavor ratio is expected to be, at the source,  $\phi^0_{\nu_e}:\phi^0_{\nu_\mu}:\phi^0_{\nu_\tau}$=1 : 2 : 0 and on Earth (due to neutrino oscillations between the source and Earth) $\phi^0_{\nu_e}:\phi^0_{\nu_\mu}:\phi^0_{\nu_\tau}$=1 : 1 : 1 and  $\phi^0_{\nu_e}:\phi^0_{\nu_\mu}:\phi^0_{\nu_\tau}$=1 : 1.8 : 1.8  for neutrino energies less and greater than 100 TeV, respectively, for gamma ray bursts ($\phi^0_{\nu_l}$ is the sum of $\nu_l$ and $\bar{\nu}_l$) \citep{kas05}.   Also it has been pointed out that measurements of the deviation of this standard flavor ratio of astrophysical high-energy neutrinos may probe new physics \citep{lea95,ath00, kas05}. 
As it is known, neutrino properties get modified when it propagates in a medium.  Depending on their flavor, neutrinos interact via neutral and/or charged currents, for instance,   $\nu_e$ interacts with electrons via both neutral and charged currents, whereas $\nu_\nu(\nu_\tau)$ interacts only via the neutral current. This induces a coherent effect in which maximal conversion of $\nu_e$ into $\nu_\mu (\nu_\tau)$ takes place. The resonant conversion of neutrino from one flavor to another due to the medium effect is well known as the Mikheyev-Smirnov-Wolfenstein effect  \citep{wol78}.  
Resonance condition of high-energy neutrinos in  hidden jets has been studied in the literature \citep{men07,raz10, sah10}.   Recently,  \citet{2013arXiv1304.4906O} studied the three-flavor neutrino oscillations on the surface of the star  for neutrino energy  in the range  (0.1 - 100) TeV.  They found that  those neutrinos generated on the surface  with  energies of less than 10 TeV could oscillate.    Unlike previous studies, we show that these sources  are capable of generating  PeV neutrinos pointing them out as possible progenitors of the first observation of PeV-energy neutrinos with IceCube \citep{aar13}. Besides,  we  do a full analysis of resonance conditions (two- and three-flavors) for neutrinos produced at different places in the star,  estimating the flavor ratios on Earth.\\
In this paper we both show that PeV neutrinos can be produced in hidden jets and  estimate the flavor ratio of high-energy neutrinos expected on Earth.  Firstly, we compute the energy range of neutrinos produced by cooling down of hadrons and mesons accelerated in a mildly relativistic jet. After that we take different matter density profiles to show that neutrinos may oscillate resonantly depending on the neutrino energy and  mixing neutrino parameters.    Finally, we discuss our results in the fail jet framework.
\section{Jet dynamics}
For the internal shocks, we consider a mildly relativistic shock propagating with bulk Lorentz factor $\Gamma_b=10^{0.5}\Gamma_{b,0.5}$.  Behind the shock, the comoving number density of particles   and density of energy are  $n'_e=n'_p=1/(8\,\pi\,m_p\,c^5)\,\Gamma_b^{-4}\,E_j\,t^{-2}_{\nu,s}\,t^{-1}_j=3.1\times10^{18}$ cm$^{-3}\,\,t^{-2}_{\nu,s}$ and $n'_p m_p c^2$, respectively, where we have taken the set of typical values for which the jet drills but hardly breaks through the stellar envelope:  the jet kinetic energy $E_j=10^{51.5} E_{j,51.5}$ erg, the variability time scale of the central object $t_\nu=t_{\nu, {\rm s}}\,{\rm s}$  with $t_{\nu,{\rm s}}$= 0.1 and 0.01, and  the jet duration $t_j=10\,t_{j,1}$ s \citep{raz05,and05,2013MNRAS.432..857M}.   We assume that electrons and protons  are accelerated in the internal shocks to a power-law distribution $N(\gamma_j) d\gamma_j\propto \gamma_j^{-p} d\gamma_j$.   The internal shocks due to shell collisions take place at a radium $r_j=2\Gamma_b^2\,c\,t_\nu= 6 \times   10^{11}\,\rm{cm}\,\Gamma^2_{0.5}\,t_{v,s}$.  Electrons, with minimum energy $E_{e,m}=\frac{p-2}{p-1} \epsilon_e\,m_p c^2 \Gamma_b$   and maximum energy limited by the dynamic time scale $t'_{dyn}\simeq t_\nu\Gamma_b$, cool down rapidly by synchrotron radiation in the presence of the magnetic field given by
\bary
B'&=&\biggl(\frac{\epsilon_B}{c^3}\,\Gamma_b^{-4}\,E_j\,t^{-2}_\nu\,t^{-1}_j  \biggr)^{1/2}\cr
&=& 3.43\times10^8\,{\rm G}\, \Gamma_{b,0.5}^{-2}\,E^{1/2}_{j,51.5}\,t^{-1/2}_{j,1}\,\epsilon_B^{1/2}\,t^{-1}_{\nu,s}\,,
\label{mfield}
\eary
where here and further on  the magnetic equipartition parameter and $t_{\nu,{\rm s}}$  lie in the range  0.1$\leq \epsilon_B\leq 10^{-4}$ and 0.1 $\leq t_{\nu,{\rm s}} \leq $  0.01, respectively. The radiated  photon energies by electron synchrotron emission  with energy $E_e$ is $E_{syn,\gamma}=eB'/(\hbar  m_e^3c^5) E^2_e$, and  also  the opacity to Thomson scattering by these photons is
\bary
\tau_{th}'&=&\frac{\sigma_T}{4\pi\,m_p\,c^4} \Gamma_b^{-3}\,E_j\,t^{-1}_\nu\,t^{-1}_j\cr
&=&3.9\times 10^5\, \Gamma_{b,0.5}^{-3}\,E_{j,51.5}\,t^{-1}_{j,1}\,t^{-1}_{\nu,s}\,.
\label{opde}
\eary
Due to the large Thomson optical depth, synchrotron photons will thermalize to a black body temperature, therefore the peak energy is given by
\bary
E'_{\gamma}\sim k_B\,T_{\gamma}&=&\biggl(\frac{15(\hbar\,c)^3}{8\pi^4\,c^3}\biggr)^{1/4}\,\epsilon_e^{1/4}\, E_j^{1/4}\,\Gamma^{-1}_b\,t_v^{-1/2}\,t^{-1/4}_j\cr
&=&1.36\,{\rm keV}\, E_{j,51.5}^{1/4}\,\Gamma^{-1}_{b,0.5}\,t^{-1/4}_{j,1}\,\epsilon^{1/4}_{e,-1}\,t_{\nu,s}^{-1/2}\,,
\label{enph}
\eary
and the number density of thermalized photons is  
\bary
\eta'_\gamma&=&\frac{2\,\zeta(3)}{\pi^2\,(c\,\hbar)^3}\,\biggl(\frac{15\,\hbar\,\epsilon_e\, E_j}{8\pi^4\,\Gamma^{4}_b\,t_v^{2}\,t_j}  \biggr)^{3/4}\cr
&=&2.86 \times 10^{23} {\rm cm^{-3}}\, E^{3/4}_{j,51.5}\,\Gamma^{-3}_{b,0.5}\,t^{-3/4}_{j,1}\,\epsilon^{3/4}_{e,-1}\,t_{\nu,s}^{-3/2}\,.
\label{denph}
\eary
Although keV photons can hardly escape due to the high optical depth,  they are  able to interact with relativistic protons accelerated in the jet, producing high-energy neutrinos via  charged pion decay. The pion energies  depend on the proton energy and  characteristics of the jet. 
\section{Hadronic model}
Protons accelerated in internal shocks, on the one hand,  radiate photons by  synchrotron radiation and  also scatter  the internal photons by  inverse Compton (IC) scattering, and on the other hand, interact with  thermal keV  photons and  hadrons by  p$\gamma$ and  p-hadron interactions.  The  optical depths for p$\gamma$  and  p-hadron  interactions are
\bary
\tau'_{p\gamma}&=&\frac{4\,\zeta(3)\sigma_{p\gamma}}{\pi^2\,(c\,\hbar)^3}\,\biggl(\frac{15\,\hbar\,\epsilon_e\, E_j}{8\pi^4\,\Gamma^{8/3}_b\,t_v^{2/3}\,t_j}  \biggr)^{3/4}\cr
&=&3.19\times 10^6\, E^{3/4}_{j,51.5}\,\Gamma^{-2}_{b,0.5}\,t^{-3/4}_{j,-1}\,\epsilon^{3/4}_{e,-1}\,t_{\nu,s}^{-1/2}\,,
\label{optpg}
\eary
and 
\bary
\tau'_{pp}&=&\frac{\sigma_{pp}}{4\,\pi\,m_p\,c^5}\,  E_j\, \Gamma^{-3}_b\,t_v^{-1}\,t_j^{-1}  \cr
&=&1.77\times 10^4\, E_{j,51.5}\, \Gamma^{-3}_{b,0.5}\,t_{j,1}^{-1}\,t_{\nu,s}^{-1}\,,
\label{optpp}
\eary
respectively. Due to the optical depths  for p$\gamma$ and p-hadron interactions are very high,  p$\gamma$ and p-hadron are effective, although p-hadron interactions are more effective at lower energy than p$\gamma$ interactions \citep{raz04b}.
\subsection{Cooling time scales}
The shock acceleration time for an energy  proton, $E'_p$, is
\bary
t'_{acc}&=&\frac{2\pi\xi}{c} r_L =\frac{2\pi\xi\,c^{1/2}\,B'_{c,p}}{m_p^2}\,E'_p\,\epsilon^{-1/2}_B\, E^{-1/2}_j\,\Gamma^{2}_b\,t_v\,t^{1/2}_j\cr
&=&2.04\times 10^{-12}{\rm s}\,E'_p\,\xi\, E^{-1/2}_{j,51.5}\,\Gamma^{2}_{b,0.5}\,t^{1/2}_{j,1}\,\epsilon^{-1/2}_B\,t_{\nu,s}\,,
\label{tacc}
\eary
where $r_L$ is the Larmor's radius and $\xi$ is a factor of equality. The acceleration time, $t'_{acc}$, gives an account of the maximum  proton energy achieved, when it is  compared with the maximum cooling time scales.  In the following subsections we are going to calculate the cooling time scales for protons and mesons.

\subsubsection{Proton cooling time scales}

The cooling time scale for proton synchrotron radiation is
\bary
t'_{p,syn}&=&\frac{E'_p}{(dE'_p/dt)_{syn}}=\frac{6\pi\,m_p^4\,c^6}{\sigma_T\,\beta^2\,m_e^2\,E'_p}\,\epsilon^{-1}_B\, E^{-1}_j\,\Gamma^{4}_b\,t^2_v\,t_j\cr
&=& 38.3\, {\rm s}\,E'^{-1}_{p,9}\, E^{-1}_{j,51.5}\,\Gamma^{4}_{b,0.5}\,t_{j,1}\,\epsilon^{-1}_B\,t^2_{\nu,s}\,.
\label{tsyn}
\eary
Protons  in the shock region  can upscatter the thermal keV photons $E'_{IC,\gamma}\sim\gamma^2_p\,E'_\gamma$  with peak energy and density given in eqs.(\ref{enph}) and (\ref{denph}).  The IC cooling time scale in the Thomson regimen is
\bary
t'^{th}_{p,ic}&=&\frac{E'_p}{(dE'_p/dt)^{th}_{ic}}=\frac{m_p^4\,c^4\,\pi^6(c\,\hbar)^2}{5\,\sigma_T\,\beta^2\,m_e^2\,\zeta(3)\,E'_p}\,\epsilon^{-1}_e\, E^{-1}_j\,\Gamma^{4}_b\,t^2_v\,t_j\cr
&=& 383.1\, {\rm s}\,E'^{-1}_{p,9}\, E^{-1}_{j,51.5}\,\Gamma^{4}_{b,0.5}\,t_{j,1}\,\epsilon^{-1}_{e,-1}\,t^2_{\nu,s}\, .
\label{tic}
\eary
Also, the IC cooling time scale in the Klein-Nishina (KN) regimen,   $E'_pE'_\gamma/m_p^2c^4=\Gamma_{KN}$ with ($\Gamma_{KN}=1$),  is
\bary
t'^{KN}_{p,ic}&=&\frac{E'_p}{(dE'_p/dt)^{KN}_{ic}}=\frac{3\pi^4(c\,\hbar)^3 \,E'_p\,\epsilon^{-1/2}_e\, E^{-1/2}_j\,\Gamma^{2}_b\,t_v\,t^{1/2}_j }{2\sqrt{30\hbar}\,\sigma_T\,\beta^2\,m_e^2\,c^5\,\zeta(3)}\cr
&=&5.15 \times 10^{-10} \, {\rm s}\,E'_{p,9}\,E^{-1/2}_{j,51.5}\,\Gamma^{2}_{b,0.5}\,t^{1/2}_{j,1}\,\epsilon^{-1/2}_{e,-1}\,t_{\nu,s}\,.
\eary
On the other hand,  protons could upscatter thermal photons according  to Bethe-Heitler (BH)  process.  The proton energy loss is taken away by the pairs produced in this process.   The cooling time scale for the BH scattering is 
\bary
t'_{BH}&=&\frac{E'_p}{(dE'_p/dt)_{BH}}=\frac{E'_p}{n'\,c\sigma_{BH}\Delta E'_p}\cr
&=&\frac{E'_p\,(m^2_p\,c^4+2E'_pE'_\gamma)^{1/2}}{2n'_\gamma\,m_e\,c^3\sigma_{BH}(E'_p+E'_\gamma)}\,,
\eary
where $\sigma_{BH}=\alpha r^2_e\, ((28/9) \,ln[2E'_p\,E'_\gamma/(m_pm_ec^4)]-106/9)$.
The energy loss rate due to pion production for p$\gamma$ interactions is \citep{ste68,bec09}
\bary
t'_{p\gamma}&=&\frac{\pi^2\,(c\,\hbar)^3}{0.3\,c\,\sigma_{p\gamma}\,\zeta(3)}\,\biggl(\frac{8\pi^4\,\Gamma^{4}_b\,t_v^{2}\,t_j}{15\,\hbar\,\epsilon_e\, E_j}  \biggr)^{3/4}\cr
&=&1.32\times10^{-5}\,\,{\rm s}\, E^{-3/4}_{j,51.5}\Gamma^{3}_{b,0.5}\,t^{3/4}_{j,1} \,\epsilon^{-3/4}_{e,-1}\,t_{\nu,s}^{3/2}\,,
\eary
and for p-hadron interactions  is  \citep{der03,der09}
\bary
t'_{pp}&=&\frac{10\,\pi\,m_p\,c^4}{\sigma_{pp}}\,  E_j^{-1}\, \Gamma^{4}_b\,t_v^{2}\,t_j \cr 
&=&4.47\times 10^{-4}\,{\rm s}\,E_{j,51.5}^{-1}\, \Gamma^{4}_{b,0.5}\,t_{j,1} \,t_{\nu,s}^{2}\,.
\eary
In figs \ref{ptime_r1}  and \ref{ptime_r2}  we have plotted the proton cooling time scales  when the magnetic field is distributed in order to 0.1$\leq \epsilon_B\leq 10^{-4}$ and internal shocks take place at  r=$6\times 10^{9}$ cm and   r=$6\times 10^{10}$ cm, respectively.
\subsubsection{Meson cooling time scales}
High-energy charged pions and kaons produced by p-hadron and p$\gamma$  interactions    ($p+\gamma/p \to X+\pi^{\pm}/K^{\pm}$)  radiate  in the presence of the magnetic field  (eq. \ref{mfield}). Therefore,  their cooling time scales  are  
\bary
t'_{\pi^+, syn}&=&  \frac{E'_{\pi^+}}{(dE'_{\pi^+}/dt)} \simeq  \frac{6\pi c^6 m^4_{\pi^+}}{\sigma_T\,\beta^2\,m_e^2}\,\epsilon^{-1}_B\, E^{-1}_j\,\Gamma^{2}_b\,t^2_v\,t_j \,E'^{-1}_{\pi^+}\cr
&=&1.9\times 10^{-2} \,{\rm s}\, E'^{-1}_{\pi^+,9}\,E^{-1}_{j,51.5}\,\Gamma^{2}_{b,0.5}\,t_{j,1}\,\epsilon^{-1}_B\,t^2_{\nu,s}\,,
\eary
and
\bary
t'_{K^+, syn}&=&  \frac{E'_{k^+}}{(dE'_{k^+}/dt)} \simeq  \frac{6\pi c^6 m^4_{k^+}}{\sigma_T\,\beta^2\,m_e^2}\,\epsilon^{-1}_B\, E^{-1}_j\,\Gamma^{2}_b\,t^2_v\,t_j  \,E'^{-1}_{k^+}\cr
&=&2.94 \,{\rm s}\,E'^{-1}_{k^+,9}\, E^{-1}_{j,51.5}\,\Gamma^{2}_{b,0.5}\,t_{j,1}\,\epsilon^{-1}_B\,t^2_{\nu,s}\,.
\eary
As protons can collide with secondary pions and kaons ($\pi^+ p$ and $K^+p$), then its respective cooling time scale is given by
\bary
t'_{had}&=&\frac{10\,\pi\,m_p\,c^4}{\sigma_{(pK/p\pi^+)}}\,  E_j^{-1}\, \Gamma^{4}_b\,t_v^{2}\,t_j \cr 
&=& 4.47\times 10^{-9} \,{\rm s}\,   E_{j,51.5}^{-1}\, \Gamma^{4}_{b,0.5}\,t_{j,1}\,t_{\nu,s}^{2}\,.
\eary
Here we have used the cross-section   $\sigma_{(pK^+/p\pi^+)} \approx 3\times 10^{-26}$ cm$^2$.  Because the mean lifetime of these mesons may be  comparable with the synchrotron  and hadron time scales in some energy range, it is necessary to consider  the cooling time  scales related to their mean lifetime which are given by
\bary
t'_{\pi^+,dec}&=&\frac{E'_{\pi^+}}{m_{\pi^+}c^2}\,\tau_{\pi^+}\cr
&=&=1.87\times 10^{-7} \,{\rm s}\,E'_{\pi^+,9}\,,
\eary
and
\bary
t'_{K^+,dec}&=&\frac{E'_{K^+}}{m_{K^+}c^2}\tau_{K^+}\cr
&=&2.51\times 10^{-8} \,{\rm s}\,E'_{K^+,9}\,,
\eary
where $\tau_{\pi^+/K^+}$is the mean lifetime for $\pi^+/K^+$ and E$_{\pi^+/K^+,9}$= 10$^9$ E$_{\pi^+/K^+}$ eV.\\
In figs \ref{mtime_r1}  and \ref{mtime_r2}  we have plotted the meson cooling time scales  when internal shocks happen at  r=$6\times 10^{9}$ cm and   r=$6\times 10^{10}$ cm and the magnetic equipartition parameter is in the range  0.1$\leq \epsilon_B\leq 10^{-4}$.
\subsection{Neutrino production}
\begin{figure}
\vspace{0.5cm}
{ \centering
\resizebox*{0.5\textwidth}{0.24\textheight}
{\includegraphics{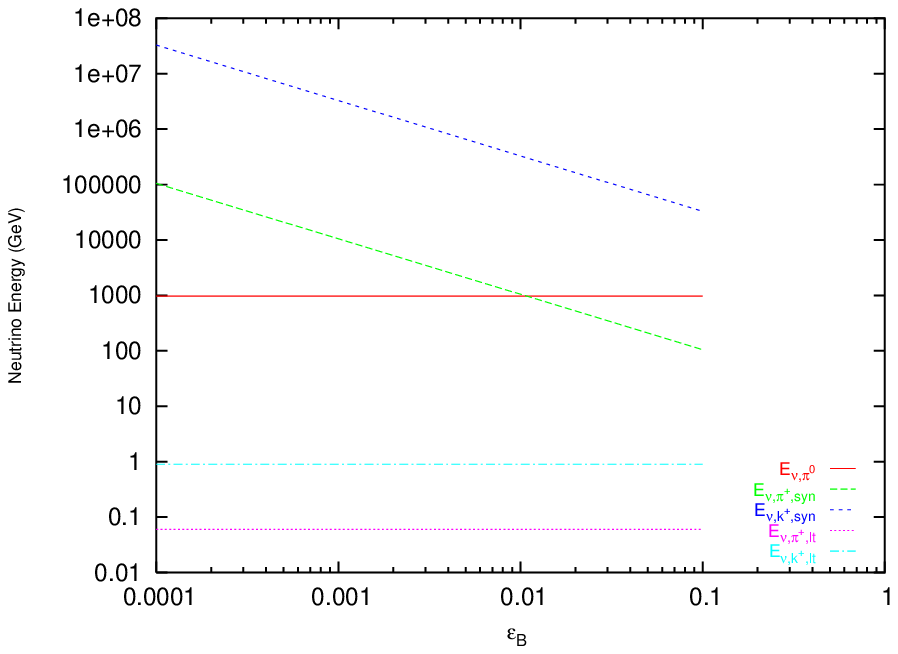}} }
{ \centering
\resizebox*{0.5\textwidth}{0.24\textheight}
{\includegraphics{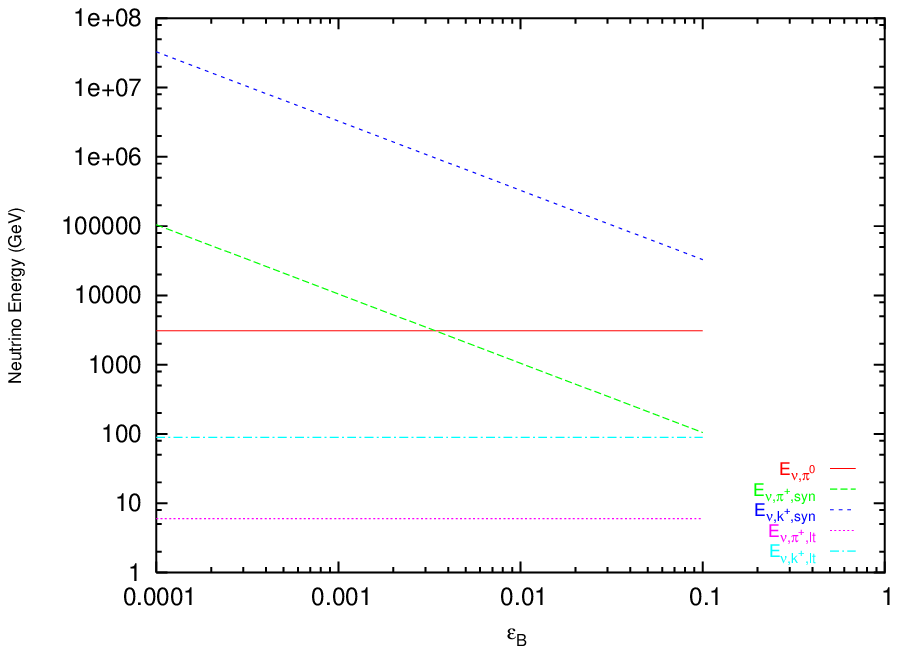}}
}
\caption{neutrino energy created at  $6\times 10^{9}$ cm (above) and $6\times 10^{10}$ cm (below)  for different interaction processes as a function of magnetic equipartition parameter.}
\label{prod_neu}
\end{figure}

The single-pion production channels are $p+\gamma\to n+\pi^+$ and $p+\gamma\to p+ \pi^0$, where the relevant pion decay chains are $\pi^0\to 2\gamma$, $\pi^+\to \mu^++\nu_\mu\to e^++\nu_e+\bar{\nu}_\mu+\nu_\mu$ and $\pi^-\to \mu^-+\bar{\nu}_\mu\to e^-+\bar{\nu}_e+\nu_\mu+\bar{\nu}_\mu$ \citep{der03}, then the threshold neutrino energy from p$\gamma$ interaction is  
\bary
E'_{\nu,\pi}&=&2.5\times 10^{-2}\biggl(\frac{8\pi^4}{15\hbar}\biggr)^{1/4}\,\,\frac{(m^2_\Delta-m_p^2)}{(1-cos\theta)}\cr
&&\hspace{3.4cm}\times\, \epsilon^{-1/4}_e\, E^{-1/4}_j\,\Gamma_b\,t^{1/2}_v\,t^{1/4}_j\cr
&=& 9.72 \,{\rm TeV}\, E^{-1/4}_{j,51.5}\,\Gamma_{b,0.5}\,t^{1/4}_{j,1}\,   \epsilon^{-1/4}_{e,-1}\,t^{1/2}_{\nu,s}\,.
\eary
Comparing the time cooling scales  we can estimate the neutrino break energy for each process. Equaling  $t_{acc}\simeq t'_{p,syn}$, we can  approximately estimate the maximum proton energy
\bary
E'_{p,max}&=&\biggl(\frac{3\,e\,m_p^4\,c^{11/2}}{\sigma_T\,\xi\,\beta^2\,m_e^2}\biggr)^{1/2}\,\epsilon^{-1/4}_B\, E^{-1/4}_j\,\Gamma_b\,t^{1/2}_v\,t^{1/4}_j\cr
&=&4.3\times 10^{3} \,{\rm GeV}\, E^{-1/4}_{j,51.5}\,\Gamma_{b,0.5}\,t^{1/4}_{j,1}\,\epsilon^{-1/4}_B\,t^{1/2}_{\nu,s}\,.
\eary
From the condition of the synchrotron cooling time scales  for mesons ($t'_{\pi^+,syn}=t'_{had}$ and $t'_{K^+,syn}=t'_{had}$), one may roughly define the neutrino break energies  as
\bary
E'_{\nu,\pi^+syn}&=&0.15\times \frac{m^4_{\pi^+}\,c^2\,\sigma_{pp}}{m_p\,\sigma_T\,\beta^2\,m_e^2}\epsilon^{-1}_B\cr
&=&10.5 \,{\rm GeV}\,  \epsilon^{-1}_{B}\,,
\eary
and
\bary
E'_{\nu,k^+syn}&=&0.3\times \frac{m^4_{k^+}\,c^2\,\sigma_{pp}}{m_p\,\sigma_T\,\beta^2\,m_e^2}\epsilon^{-1}_B\cr
&=&3.28 \,{\rm TeV}\,\epsilon^{-1}_{B}\,.
\eary
From the lifetime condition of  cooling time scale  ($t'_{\pi^+,dec}=t'_{had}$ and  $t'_{K^+,dec}=t'_{had}$),  one again we can obtain  the neutrino break energies, which for these cases are 
\bary
E'_{\nu, \pi^+lt}&=&2.5\frac{\pi\,m_p\,m_{\pi^+}\,c^6}{\sigma_{pp}}\,\tau^{-1}_{\pi^+}  \,E^{-1}_j\,\Gamma^{4}_b\,t^2_v\,t_j\cr
&=&0.6 \,{\rm TeV}\,E^{-1}_{j,51.5}\,\Gamma^{4}_{b,0.5}\,t_{j,1}\,t^2_{\nu,s}\,,
\eary
and
\bary
E'_{\nu, k^+lt}&=&5\frac{\pi\,m_p\,m_{k^+}\,c^6}{\sigma_{pp}}\,\tau^{-1}_{k^+}  \,E^{-1}_j\,\Gamma^{4}_b\,t^2_v\,t_j\cr
&=&8.92 \,{\rm TeV}\,E^{-1}_{j,51.5}\,\Gamma^{4}_{b,0.5}\,t_{j,1}\,t^2_{\nu,s}\,.
\eary
It is important to say that muons may be suppressed by electromagnetic energy losses and in that case would not contribute  much to high-energy neutrino production. The ratio $\frac{t'_{\pi^+/K^+,cool}}{t'_{\pi^+/K^+,dec}}$, where $t'_{\pi^+/K^+,cool}=\frac{t'_{\pi^+/K^+,em}\cdot \,\,\,t'_{\pi^+/K^+,had}} {t'_{\pi^+/K^+,em}\,\,+\, \,\,t'_{\pi^+/K^+,had} }$, determines the suppression of mesons before they decay to neutrinos \citep{raz05}. \\
In fig. \ref{prod_neu} we have plotted the neutrino energy created by distinct interaction processes at  different distances,   $6\times 10^{9}$ cm (above) and $6\times 10^{10}$ cm (below),  as a function of the magnetic equipartition parameter.
\section{Density profile of the source}
Analytical and numerical models of density distribution in a  pre-supernova have shown a decreasing dependence  on radius  $\rho\propto r^{-n}$, with n=3/2 - 3 above the core, being 3/2 and 3  convective and radiative envelopes respectively \citep{woo93, shi90, arn91}.  In particular,  distributions with $\rho \propto r^{-3}$ and $\rho \propto r^{-17/7}$ have been proposed  to describe simple blast wave distributions \citep{bet90, che89}.   Following \cite{men07}, we use three models of  density profile; Model [A], Model [B] and Model [C].\\ Model [A] ,
\be
\hspace{0.7cm}{\rm [A]} ~~\rho(r) = 4.0\times 10^{-6} \left( \frac{R_\star}{r} -1\right)^3 ~{\rm g~cm}^{-3}\,,
\label{dens-pro-A} \\
\ee
corresponds to a polytropic hydrogen envelope with $\rho(r)\propto r^{-3}$, scaling valid in the range $r_{jet}\geq r \geq R_\star$.    Model [B], 
\bary
&&{\rm [B]}~~\rho(r) = 3.4\times 10^{-5} ~{\rm g~cm}^{-3}\cr
&&\times
\cases{ 
(R_\star/r)^{17/7}\,; \hspace{1cm}10^{10.8} ~{\rm cm}<r<r_b = 10^{12} ~{\rm cm}   &\nonumber \cr
(R_\star/r_b)^{17/7} (r-R_\star)^5/(r_b - R_\star)^5\,;\hspace{0.3cm} r>r_b\,,& \cr
} 
\label{dens-pro-B} \\
\eary
is a power-law fit with an effective polytropic index $n_{eff}=17/7$ as done for SN 1987A \citep{che89}.  Here $r_j \sim  10^{10.8} ~{\rm cm} $ is the radius of inner border of the envelope, where the density $\rho=0.4$ g cm$^{-3}$. Associating the number of electron per nucleon $Y_e=$0.5, we obtained the number density of electrons as $N_e=N_a\,\rho(r)\, Y_e=$1.2$\times$10$^{23}$ cm$^{-3}$ and    Model [C], 
 \bary
{\rm [C]} ~~\rho(r) = 6.3\times 10^{-6} {\cal A}
\left( \frac{R_\star}{r} -1 \right)^{n_{\rm eff}}  
~{\rm g~cm}^{-3} \cr
(n_{\rm eff}, {\cal A})=
\cases{ 
(2.1,20) ~;~ &$10^{10.8} ~{\rm cm}< r < 10^{11} ~{\rm cm}$ \cr
(2.5,1) ~;~ &$r > 10^{11} ~{\rm cm}$\,, \cr
}
\label{dens-pro-C}
\eary
includes a sharp drop in density at the edge of the helium core \citep{mat99}. 
\section{Neutrino Mixing}
In the following subsections we are going to describe the neutrino oscillations in the matter (along the jet for three density profiles given in section 4 ) and  in vacuum (its path up to  Earth). We will be using  the best fit parameters for two-neutrino mixing (solar, atmospheric and accelerator neutrino experiments) and three-neutrino mixing.
The best fit value of   solar, atmospheric and accelerator neutrino experiments  are given as follow.\\
\textbf{Solar Neutrinos} are electron neutrinos produced in the thermonuclear reactions which generate the solar energy.  The Sudbury Neutrino Observatory (SNO) was designed to measure the flux of neutrinos produced by $^8$B decays in the sun, so-called $^8$B neutrinos, and to study neutrino oscillations, as proposed by \cite{che85}.  A two-flavor neutrino oscillation analysis gave the following parameters:  $\delta m^2=(5.6^{+1.9}_{-1.4})\times 10^{-5}\,{\rm eV^2}$ and $\tan^2\theta=0.427^{+0.033}_{-0.029}$\citep{aha11}.\\
\textbf{Atmospheric Neutrinos} are electron neutrinos  $\nu_e$  produced mainly from the decay chain $\pi\to \mu+\nu_\mu$ followed by $\mu\to e+\nu_\mu+\nu_e$.  Super-Kamiokande (SK) observatory  observes interactions between neutrinos with electrons or with nuclei or water   via the water Cherenkov method. Under a two-flavor disappearance model with separate mixing parameters between neutrinos and antineutrinos there were found the following parameters for the SK-I + II + III data: $\delta m^2=(2.1^{+0.9}_{-0.4})\times 10^{-3}\,{\rm eV^2}$ and $\sin^22\theta=1.0^{+0.00}_{-0.07}$ \citep{abe11a}.\\
 \textbf{Reactor Neutrinos} are  produced in Nuclear reactors.  Kamioka Liquid scintillator Anti-Neutrino Detector (KamLAND) was initially designed to detect reactor neutrinos and so later it was prepared to measure $^7$Be solar neutrinos. A two neutrino oscillation analysis gives  $\delta m^2=(7.9^{+0.6}_{-0.5})\times 10^{-5}\,{\rm eV^2}$ and $\tan^2\theta=0.4^{+0.10}_{-0.07}$\citep{ara05,shi07,mit11}.\footnote{this value was obtained using a global analysis of data from KamLAND and solar-neutrino experiments}.\\
\textbf{Accelerator Neutrinos} are mostly produced by $\pi$ decays (and some K decays), with the pions produced by the scattering of the accelerated protons on a fixed target. The beam can contain both $\mu$- and e-neutrinos and antineutrinos.  There are two categories: Long and short baselines.\\
Long-baseline experiments with accelerator beams run  with a baseline of about a hundred of kilometers.  K2K experiment was designed to measure neutrino oscillations using a man-made beam with well controlled systematics, complementing and confirming the measurement made with atmospheric neutrinos.  $\delta m^2=(2.8^{+0.7}_{-0.9})\times 10^{-3}\,{\rm eV^2}$ and $\sin^22\theta=1.0$\citep{ahn06}.\\
Short-baseline experiments with accelerator beams run  with a baseline of about hundreds of meters. Liquid Scintillator Neutrino Detector (LSND) was designed to search for $\nu_\mu\to\nu_e$ oscillations using $\nu_\mu$ from $\pi^+$ decay in flight \citep{ath96, ath98}.  The region of parameter space has been partly tested by Karlsruhe Rutherford medium energy neutrino KARMEN \citep{arm02} and MiniBooNe experiments. \cite{chu02} found two well-defined regions of oscillation parameters with either $\delta m^2  \approx  7\, {\rm eV^2}$ or $\delta m^2 < 1\, {\rm eV^2} $ compatible with both LAND and KARMEN experiments, for the complementary confidence. The MiniBooNE experiment was specially designed to verify the LSND's neutrino data. It  is currently running at Fermilab and is searching for  $\nu_e (\bar{\nu}_e)$ appearance in a $\nu_\mu(\bar{\nu}_\mu)$ beam. Although MiniBooNE found no evidence for an excess of $\nu_e$ candidate events above 475 MeV in the $\nu_\mu\to\nu_e$ study, there was observed a 3.0$\sigma$ excess of electron-like events below 475 MeV\citep{agu09,agu10,agu07}.  In addition, in the $\bar{\nu}_\mu\to\bar{\nu}_e$ study, MiniBooNE found evidence of oscillations in the 0.1 to 1.0 eV$^2$, which are consistent with LSND results \citep{ath96, ath98}.\\
Combining solar, atmospheric, reactor and accelerator parameters, the best fit values of the three neutrino mixing  are

 for $\sin^2_{13} < 0.053$: \citep{aha11}
 \be
 \Delta m_{21}^2= (7.41^{+0.21}_{-0.19})\times 10^{-5}\,{\rm eV^2};   \hspace{0.1cm} \tan^2\theta_{12}=0.446^{+0.030}_{-0.029}\,,
 \ee
and  for $\sin^2_{13} < 0.04$: \citep{wen10}
\be
\Delta m_{23}^2=(2.1^{+0.5}_{-0.2})\times 10^{-3}\,{\rm eV^2};    \hspace{0.1cm} \sin^2\theta_{23}=0.50^{+0.083}_{-0.093}\,
\label{3parosc}
\ee
\subsection{Neutrino oscillation inside the jet}
When  neutrino oscillations take place in the matter, a resonance could occur that dramatically enhances the flavor mixing and can lead to maximal conversion from one neutrino flavor to another.  This resonance depends on the effective potential,  density profile of the medium,  and oscillation parameters. As  $\nu_e$ is the one that can  interact  via CC,   the effective potential can be obtained  calculating the difference between the potential due to CC and  NC contributions \citep{kuo89}.
\subsubsection{Two-Neutrino Mixing}
In this subsection, we will consider the neutrino oscillation process $\nu_e\leftrightarrow \nu_{\mu, \tau}$. The evolution equation for the propagation of neutrinos in the above medium is given by
\be
i
{\pmatrix {\dot{\nu}_{e} \cr \dot{\nu}_{\mu}\cr}}
={\pmatrix
{V_{eff}-\Delta \cos 2\theta & \frac{\Delta}{2}\sin 2\theta \cr
\frac{\Delta}{2}\sin 2\theta  & 0\cr}}
{\pmatrix
{\nu_{e} \cr \nu_{\mu}\cr}},
\ee
where $\Delta=\delta m^2/2E_{\nu}$, $V_{eff}=\sqrt 2G_F\, N_e$ is the effective potential,     $E_{\nu}$ is the neutrino energy,   and $\theta$ is the neutrino mixing angle. For anti-neutrinos one has to replace $N_e$ by $-N_e$. The conversion probability for a given time $t$ is
\be
P_{\nu_e\rightarrow {\nu_{\mu}{(\nu_\tau)}}}(t) = 
\frac{\Delta^2 \sin^2 2\theta}{\omega^2}\sin^2\left (\frac{\omega t}{2}\right
),
\label{prob}
\ee
with
\be
\omega=\sqrt{(V_{eff}-\Delta \cos 2\theta)^2+\Delta^2 \sin^2
    2\theta}.
\ee
The oscillation length for the neutrino is given by
\be
L_{osc}=\frac{L_v}{\sqrt{\cos^2 2\theta (1-\frac{V_{eff}}{\Delta \cos 2\theta})^2+\sin^2 2\theta}},
\label{osclength}
\ee
where $L_v=2\pi/\Delta$ is the vacuum oscillation length. If the density of the medium is such that the condition $\sqrt2 G_F\,N_e=\Delta\,\cos2\theta$ is satisfied,   the resonant condition, 

\be
V_{eff,R}=\Delta \cos 2\theta\,,
\label{reso2d}
\ee
can come about, therefore the resonance length can be written as
\be
L_{res}=\frac{L_v}{\sin 2\theta}.
\label{oscres}
\ee
Combining eqs (\ref{oscres}) and (\ref{reso2d}) we can obtain the resonance density as a function of resonance length
{\scriptsize 
\begin{equation}\label{p1}
\textbf{$\rho_R$}=
\cases{
\frac{3.69\times 10^{-4}}{E_{\nu,TeV}}	\,     \biggl[ 1- E_{\nu,TeV}^2\biggl( \frac{4.4 \times 10^{12} \,cm}{l_r}\biggr)^2  \biggr]^{1/2}{\rm gr/cm^3}       &   {\rm sol.} \,, \nonumber\cr
\frac{1.39\times 10^{-2}}{E_{\nu,TeV}}	\,     \biggl[ 1- E_{\nu,TeV}^2\biggl( \frac{1.18 \times 10^{11}\,cm}{l_r}\biggr)^2  \biggr]^{1/2} {\rm gr/cm^3}     &   {\rm atmosp.}\,,\nonumber\cr
\frac{3.29}{E_{\nu,TeV}}	                             \,     \biggl[ 1- E_{\nu,TeV}^2\biggl( \frac{4.9 \times 10^8\,cm}{l_r}\biggr)^2  \biggr]^{1/2} {\rm gr/cm^3}           &   {\rm accel.}\,, \cr
}
\end{equation}
}
\noindent where sol, atmosp. and accel. correspond to solar, atmospheric and accelerator  parameters.\\
In addition of the resonance condition, the dynamics of this transition must be  determined by adiabatic conversion through the adiabaticity  parameter 
 \be
\gamma\equiv \frac{\delta m^2}{2E}\sin2 \theta\,\tan2 \theta\frac{1}{\mid \frac1\rho\, \frac {d\rho}{dr}\mid}_R\,,
\ee
with  $\gamma\gg$ 1  or the flip probability  given by
\be
 P_f= e^{-\pi/2\,\gamma}\,,
 \label{flip}
 \ee
 where $\rho$ is given by eqs. (\ref{dens-pro-A}), (\ref{dens-pro-B})  and (\ref{dens-pro-C}).


\subsubsection{Three-neutrino Mixing}
To determine the neutrino oscillation probabilities we have to solve the evolution equation of the neutrino system in the matter. In a three-flavor framework, this equation is given by
\be
i\frac{d\vec{\nu}}{dt}=H\vec{\nu},
\ee
and the state vector in the flavor basis is defined as
\be
\vec{\nu}\equiv(\nu_e,\nu_\mu,\nu_\tau)^T.
\ee
The effective Hamiltonian is
\be
H=U\cdot H^d_0\cdot U^\dagger+diag(V_{eff},0,0),
\ee
with
\be
H^d_0=\frac{1}{2E_\nu}diag(-\Delta m^2_{21},0,\Delta^2_{32}).
\ee
with the same potential $V_{eff}$ given for two-neutrino mixing subsection and $U$  the three
neutrino  mixing matrix given by \cite{gon03,akh04,gon08,gon11}
\be
U =
{\pmatrix
{
c_{13}c_{12}                    & s_{12}c_{13}                    & s_{13}\cr
-s_{12}c_{23}-s_{23}s_{13}c_{12} & c_{23}c_{12}-s_{23}s_{13}s_{12}   & s_{23}c_{13}\cr
s_{23}s_{12}-s_{13}c_{23}c_{12}  &-s_{23}c_{12}-s_{13}s_{12}c_{23}   &  c_{23}c_{13}\cr
}},
\ee
where $s_{ij}=\sin\theta_{ij}$ and  $c_{ij}=\cos\theta_{ij}$ and we have taken the Dirac phase $\delta=0$. For anti-neutrinos one has to replace $U$ by $U^*$.  The different neutrino probabilities are given as
%
{\scriptsize 
\bary
P_{ee}&=&1-4s^2_{13,m}c^2_{13,m}S_{31}\,,\nonumber\\
P_{\mu\mu}&=&1-4s^2_{13,m}c^2_{13,m}s^4_{23}S_{31}-4s^2_{13,m}s^2_{23}c^2_{23}S_{21}-4
c^2_{13,m}s^2_{23}c^2_{23}S_{32}\,,\nonumber\\
P_{\tau\tau}&=&1-4s^2_{13,m}c^2_{13,m}c^4_{23}S_{31}-4s^2_{13,m}s^2_{23}c^2_{23}S_{21}-4
c^2_{13,m}s^2_{23}c^2_{23}S_{32}\,,\nonumber\\
P_{e\mu}&=&4s^2_{13,m}c^2_{13,m}s^2_{23}S_{31}\,,\nonumber\\
P_{e\tau}&=&4s^2_{13,m}c^2_{13,m}c^2_{23}S_{31}\,,\nonumber\\
P_{\mu\tau}&=&-4s^2_{13,m}c^2_{13,m}s^2_{23}c^2_{23}S_{31}+4s^2_{13,m}s^2_{23}c^2_{23}S_{21}+4
c^2_{13,m}s^2_{23}c^2_{23}S_{32}\,,\nonumber\\
\eary
}
%
where
\be
\sin
2\theta_{13,m}=\frac{\sin2\theta_{13}}{\sqrt{(\cos2\theta_{13}-2E_{\nu}V_e/\delta
    m^2_{32})^2+(\sin2\theta_{13})^2}},
\ee
and
\be
S_{ij}=\sin^2\biggl(\frac{\Delta\mu^2_{ij}}{4E_{\nu}}L\biggr).
\ee
Here $\Delta\mu^2_{ij}$ are given by 
\bary
\Delta\mu^2_{21}&=&\frac{\Delta
  m^2_{32}}{2}\biggl(\frac{\sin2\theta_{13}}{\sin2\theta_{13,m}}-1\biggr)-E_{\nu}V_e\,,\nonumber\\
\Delta\mu^2_{32}&=&\frac{\Delta
  m^2_{32}}{2}\biggl(\frac{\sin2\theta_{13}}{\sin2\theta_{13,m}}+1\biggr)+E_{\nu}V_e\,,\nonumber\\
\Delta\mu^2_{31}&=&\Delta m^2_{32} \frac{\sin2\theta_{13}}{\sin2\theta_{13,m}}\,,
\eary
where
\bary
\sin^2\theta_{13,m}&=&\frac12\biggl(1-\sqrt{1-\sin^22\theta_{13,m}}\biggr)\,,\nonumber\\
\cos^2\theta_{13,m}&=&\frac12\biggl(1+\sqrt{1-\sin^22\theta_{13,m}}\biggr)\,.
\eary
The oscillation length for the neutrino is given by
\be
L_{osc}=\frac{L_v}{\sqrt{\cos^2 2\theta_{13} (1-\frac{2 E_{\nu} V_e}{\delta m^2_{32} \cos 2\theta_{13}} )^2+\sin^2 2\theta_{13}}},
\label{osclength}
\ee
where $L_v=4\pi E_{\nu}/\delta m^2_{32}$ is the vacuum oscillation length. From the resonance condition,  $\sqrt2 G_F\,N_e=\Delta \cos2\theta_{13}$, the resonance length and density  are related as
\be 
\rho_R=\frac{1.9\times 10^{-2}}{E_{\nu,TeV}}	\,     \biggl[ 1- E_{\nu,TeV}^2\biggl( \frac{8.2 \times 10^{10} \,cm}{l_r}\biggr)^2  \biggr]^{1/2}{\rm gr/cm^3}\,.
\label{p2}
\ee
On the other hand, generalizing the adiabaticity  parameter, $\gamma$, to three-mixing neutrinos, it can be written as
 \be
\gamma\equiv \frac{\delta m_{32}^2}{2E}\sin2 \theta_{13}\,\tan2 \theta_{13}\frac{1}{\mid \frac1\rho\, \frac {d\rho}{dr}\mid}_R\,,
\ee
with the flip probability given by eq.  (\ref{flip}).
\subsection{Neutrino Oscillation from  Source to Earth}
Between the surface of the star and the Earth the flavor ratio $\phi^0_{\nu_e}:\phi^0_{\nu_\mu}:\phi^0_{\nu_\tau}$   is affected by the full three description  flavor mixing, which is calculated as follow. The probability for a neutrino to oscillate from a flavor estate $\alpha$ to a flavor state $\beta$ in a time starting from the emission of neutrino at star t=0, is given as
\bary
P_{\nu_\alpha\to\nu_\beta} &=&\mid <  \nu_\beta(t) | \nu_\alpha(t=0) >  \mid\cr
&=&\delta_{\alpha\beta}-4 \sum_{j>i}\,U_{\alpha i}U_{\beta i}U_{\alpha j}U_{\beta i}\,\sin^2\biggl(\frac{\delta m^2_{ij} L}{4\, E_\nu}   \biggr)\,.
\eary
Using the set of parameters give in eq. (\ref{3parosc}), we can write the mixing matrix
\be
U =
{\pmatrix
{
0.816669	   &  0.544650     &     0.190809\cr
 -0.504583  & 0.513419      &	 0.694115\cr
 0.280085   &  -0.663141    &     0.694115\cr
}}\,.
\ee
Averaging the sin term in the probability to $\sim 0.5$ for larger distances L \citep{lea95}, the probability matrix for a neutrino flavor vector of ($\nu_e$, $\nu_\mu$, $\nu_\tau$)$_{source}$ changing to a flavor vector  ($\nu_e$, $\nu_\mu$, $\nu_\tau$)$_{Earth}$ is given as
\be
{\pmatrix
{
\nu_e   \cr
\nu_\mu   \cr
\nu_\tau   \cr
}_{Earth}}
=
{\pmatrix
{
0.534143	  & 0.265544	  & 0.200313\cr
 0.265544	  & 0.366436	  &  0.368020\cr
 0.200313	  & 0.368020	  & 0.431667\cr
}}
{\pmatrix
{
\nu_e   \cr
\nu_\mu   \cr
\nu_\tau   \cr
}_{source}}
\label{matrixosc}
\ee
for distances longer than the solar system. 
\section{Results and Discussions}
\begin{figure}
\vspace{0.5cm}
{ \centering
\resizebox*{0.45\textwidth}{0.22\textheight}
{\includegraphics{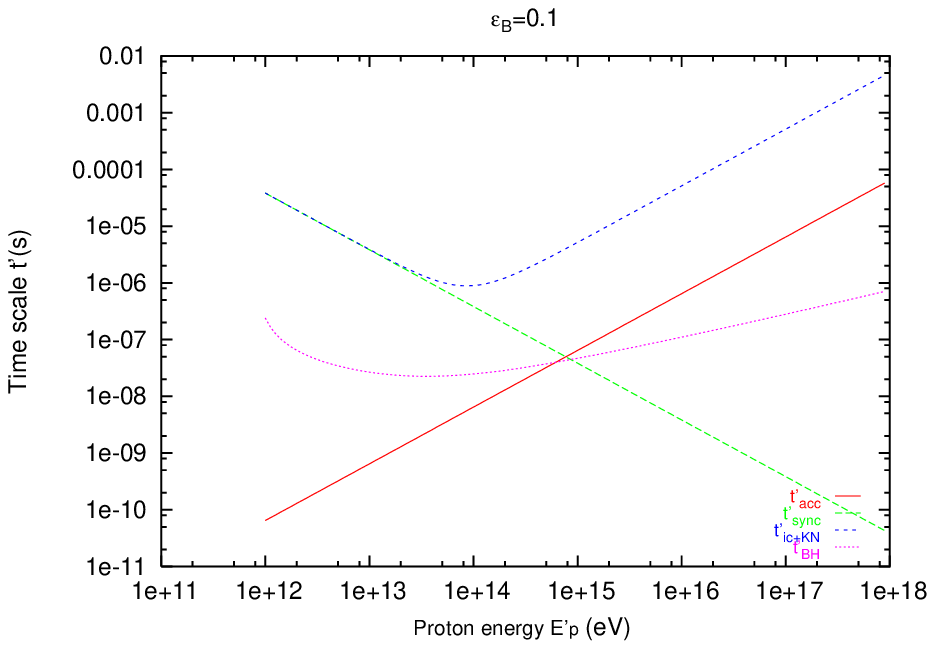}}
\resizebox*{0.45\textwidth}{0.22\textheight}
{\includegraphics{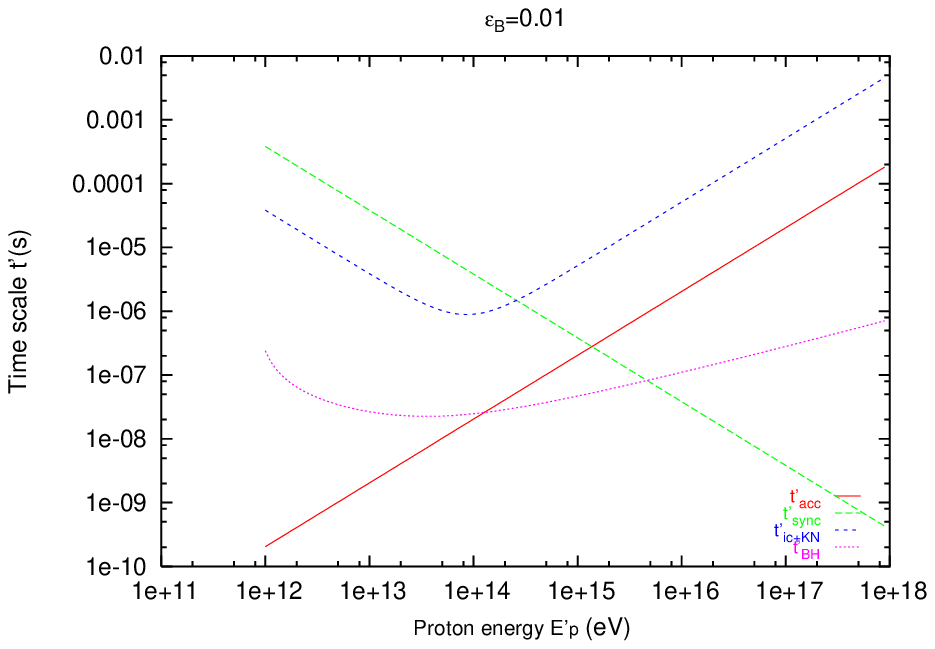}}
\resizebox*{0.45\textwidth}{0.22\textheight}
{\includegraphics{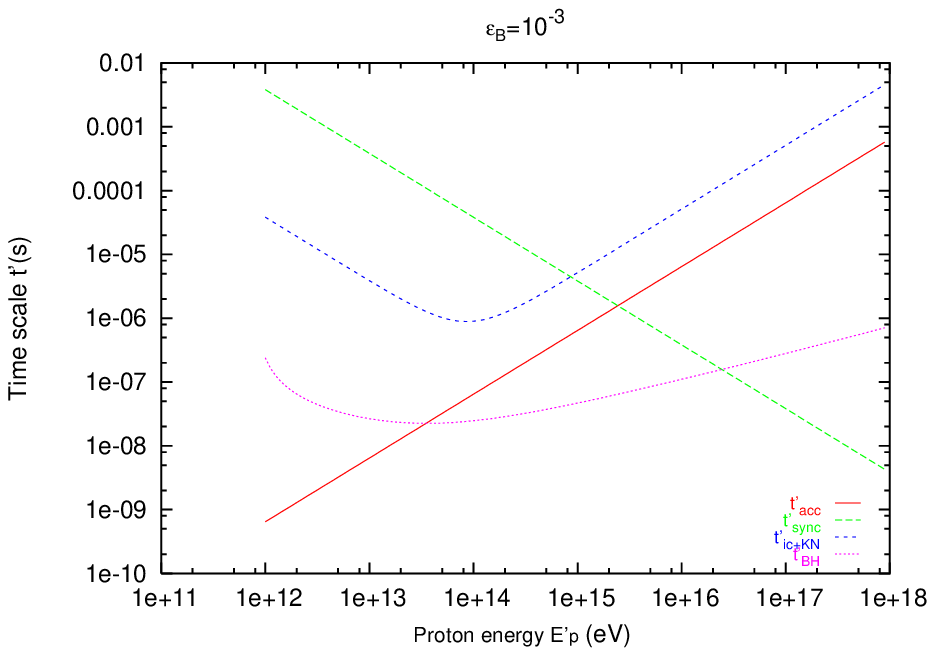}}
\resizebox*{0.45\textwidth}{0.22\textheight}
{\includegraphics{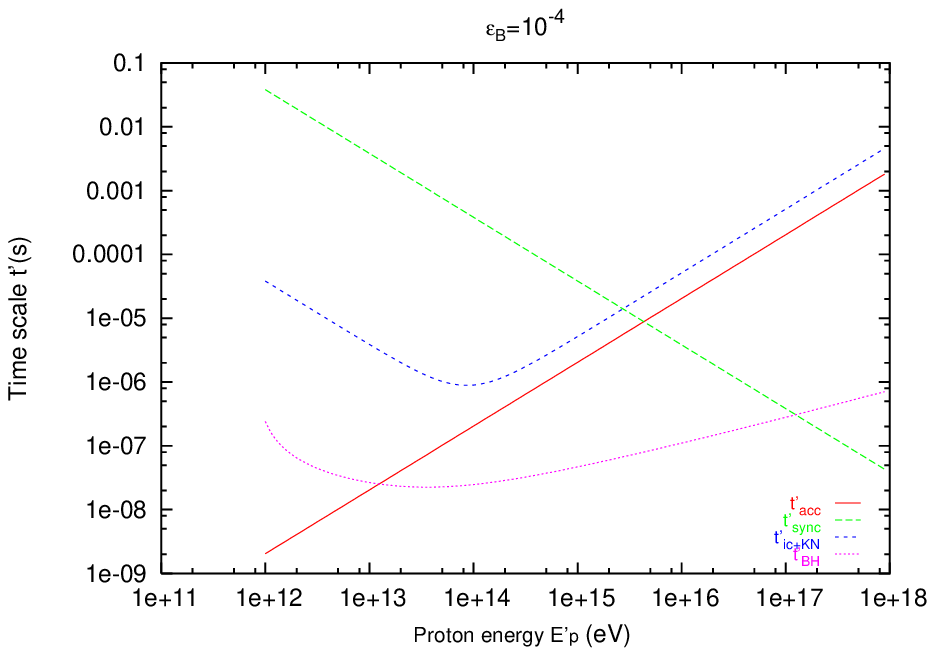}}
}
\caption{Proton cooling time scales in the comoving frame for different processes as a function of proton energy (E$_p$) when the shell collisions take place  at r=$6\times 10^9$ cm and  different magnetic fields. Synchrotron radiation (t$'_{sync}$), IC+KN scattering (t$'_{ic+KN}$),  Bethe-Heitler (t$'_{BH}$), shock acceleration time (t$'_{acc}$) }
\label{ptime_r1}
\end{figure}

\begin{figure}
\vspace{0.5cm}
{ \centering
\resizebox*{0.45\textwidth}{0.22\textheight}
{\includegraphics{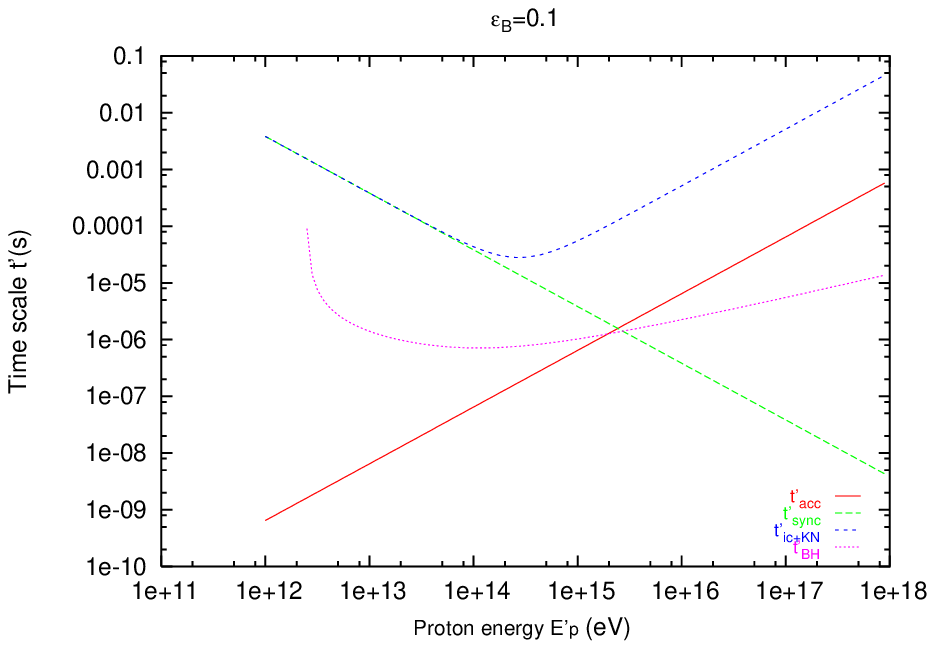}}
\resizebox*{0.45\textwidth}{0.22\textheight}
{\includegraphics{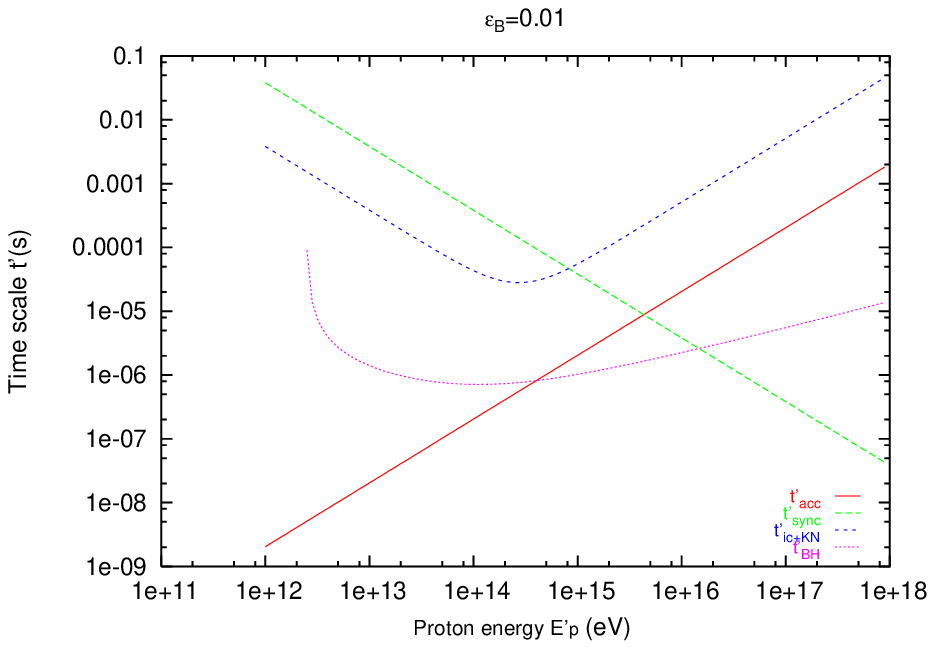}}
\resizebox*{0.45\textwidth}{0.22\textheight}
{\includegraphics{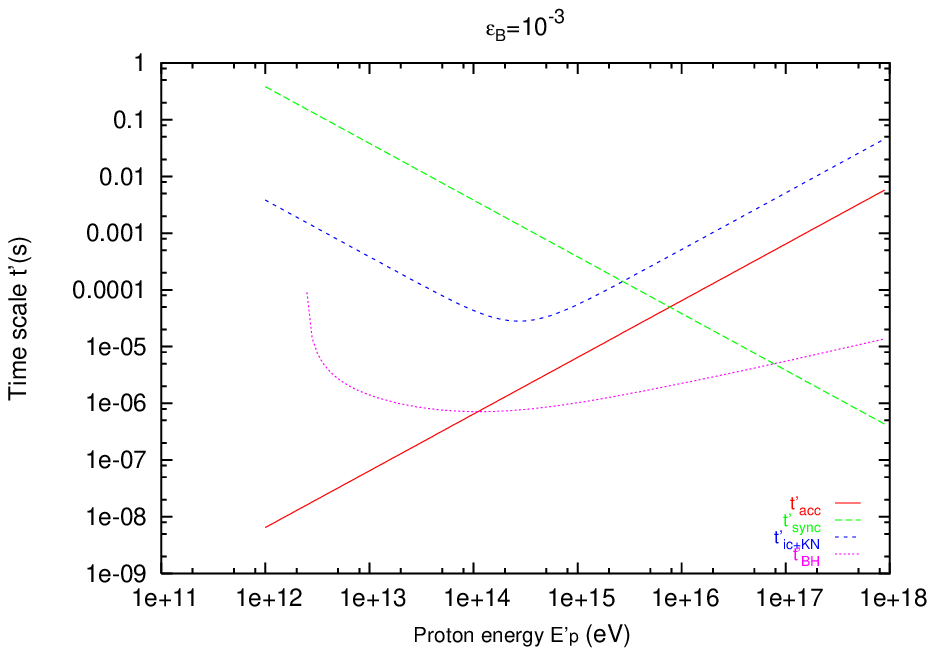}}
\resizebox*{0.45\textwidth}{0.22\textheight}
{\includegraphics{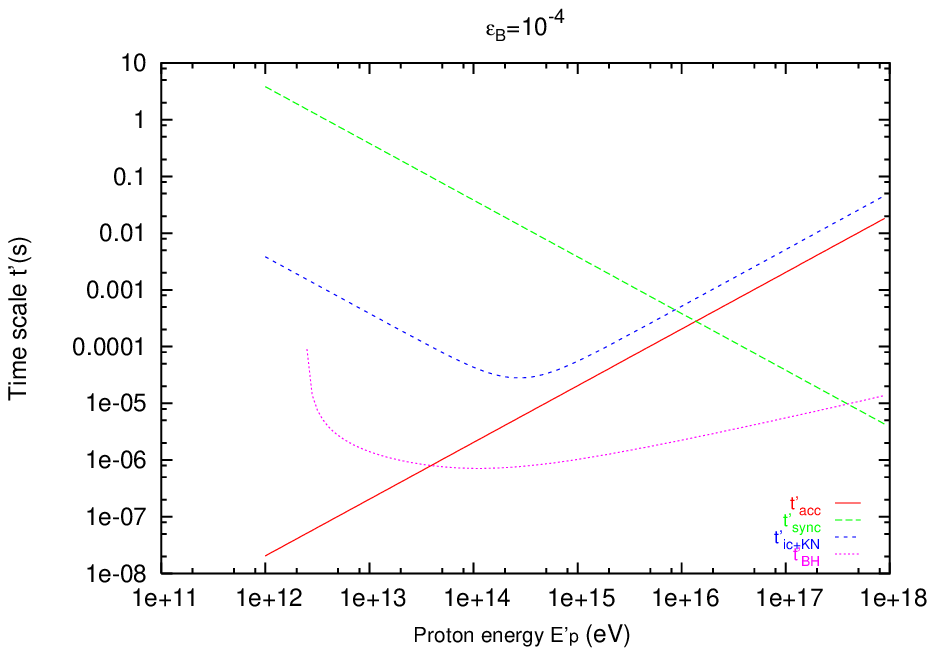}}
}
\caption{Proton cooling time scales in the comoving frame for different processes as a function of proton energy (E$_p$) when the shell collisions take place  at r=$6\times 10^{10}$ cm and  different magnetic fields.  Synchrotron radiation (t$'_{sync}$), IC+KN scattering (t$'_{ic+KN}$),  Bethe-Heitler (t$'_{BH}$), shock acceleration time (t$'_{acc}$) }
\label{ptime_r2}
\end{figure}

\begin{figure}
\vspace{0.5cm}
{ \centering
\resizebox*{0.45\textwidth}{0.22\textheight}
{\includegraphics{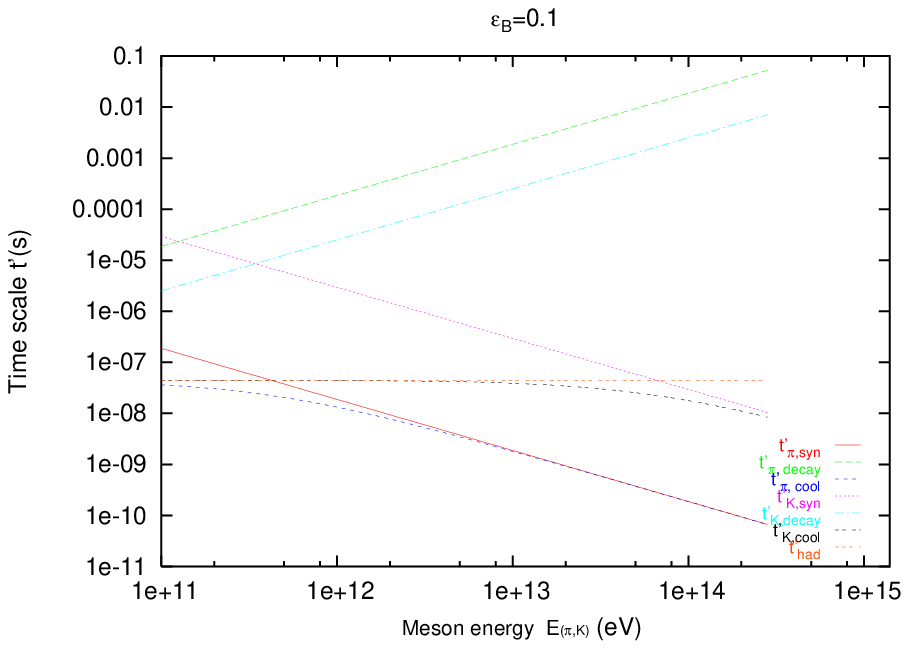}}
\resizebox*{0.45\textwidth}{0.22\textheight}
{\includegraphics{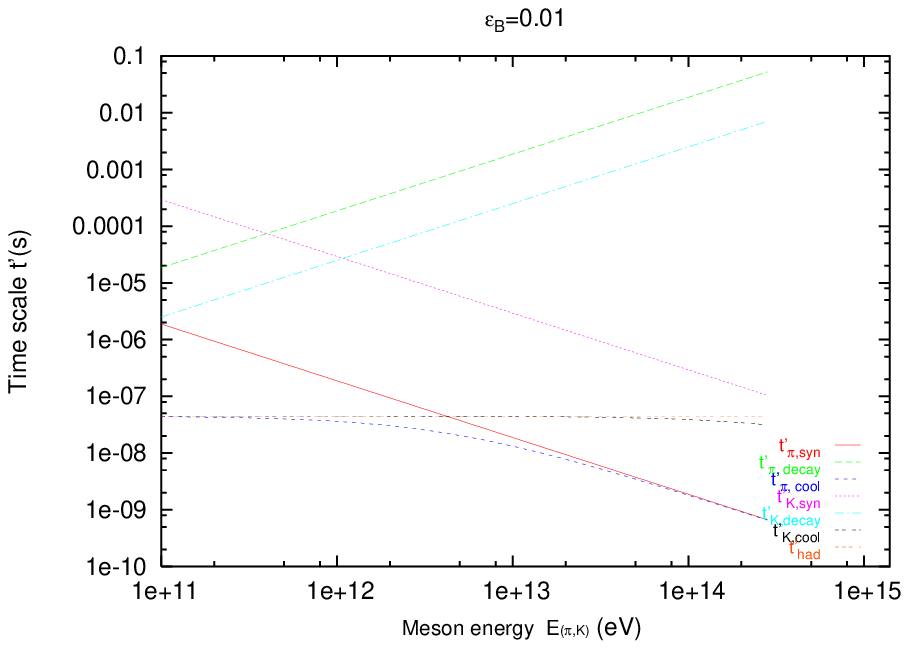}}
\resizebox*{0.45\textwidth}{0.22\textheight}
{\includegraphics{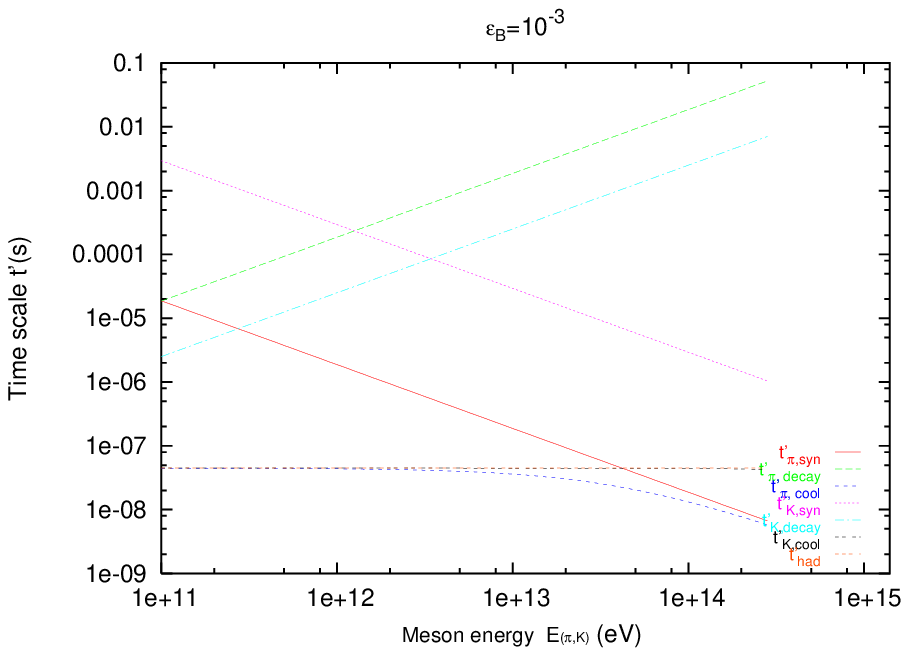}}
\resizebox*{0.45\textwidth}{0.22\textheight}
{\includegraphics{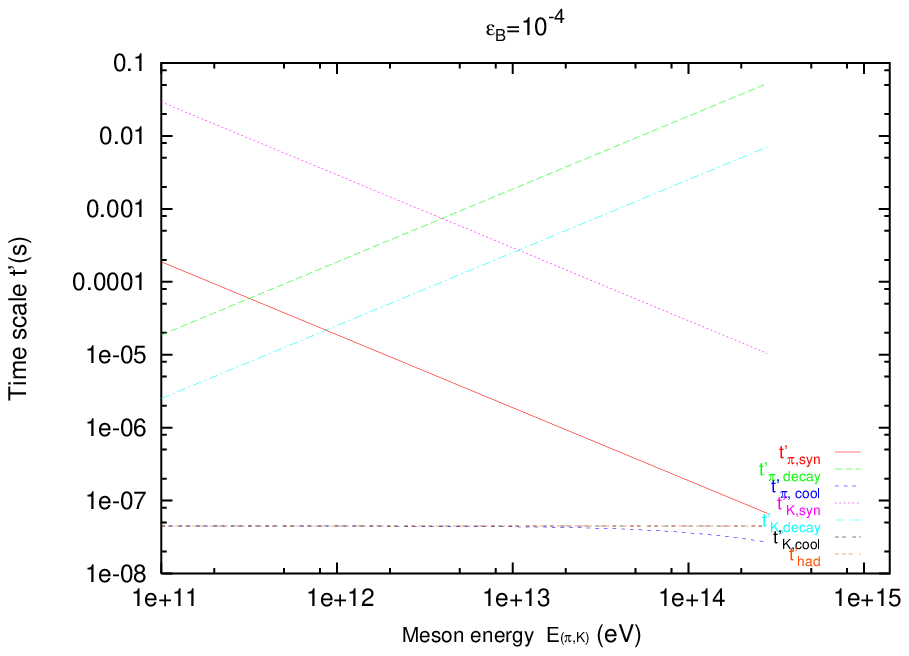}}
}
\caption{Meson $(\pi^{\pm}/K^{\pm})$ cooling time scales in the comoving frame for different processes as a function of Meson energy (E$_{\pi/K}$)  when the shell collisions take place  at r=$6\times 10^9$ cm and  different magnetic fields.  $\pi/K$- synchrotron radiation (t$'_{(\pi/K),sync}$), $\pi$/K- decay (t$'_{(\pi/K),decay}$),  $\pi$/K- synchrotron and hadronic  radiation    (t$'_{cool}$).   }
\label{mtime_r1}
\end{figure}

\begin{figure}
\vspace{0.5cm}
{ \centering
\resizebox*{0.45\textwidth}{0.22\textheight}
{\includegraphics{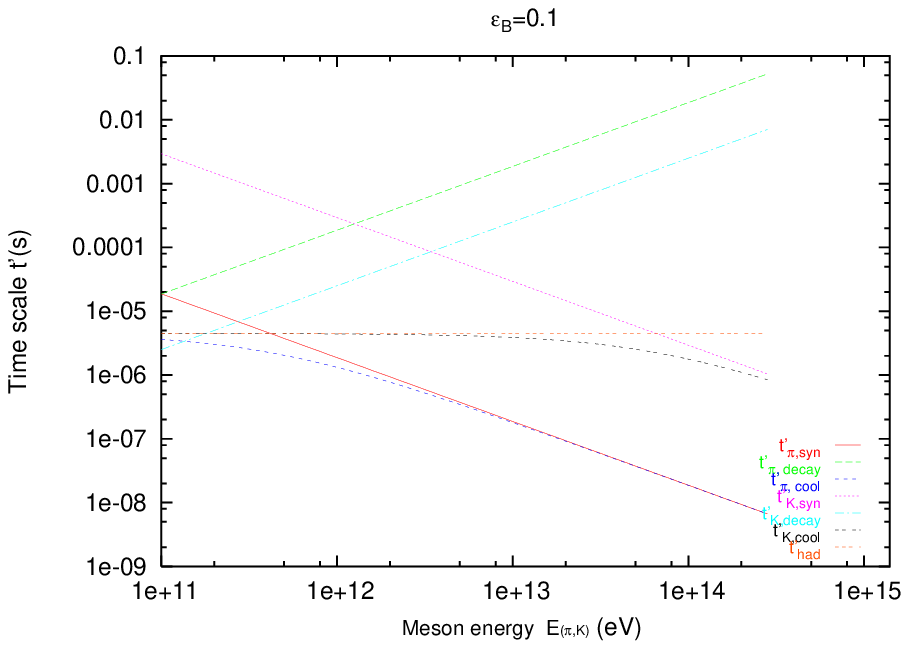}}
\resizebox*{0.45\textwidth}{0.22\textheight}
{\includegraphics{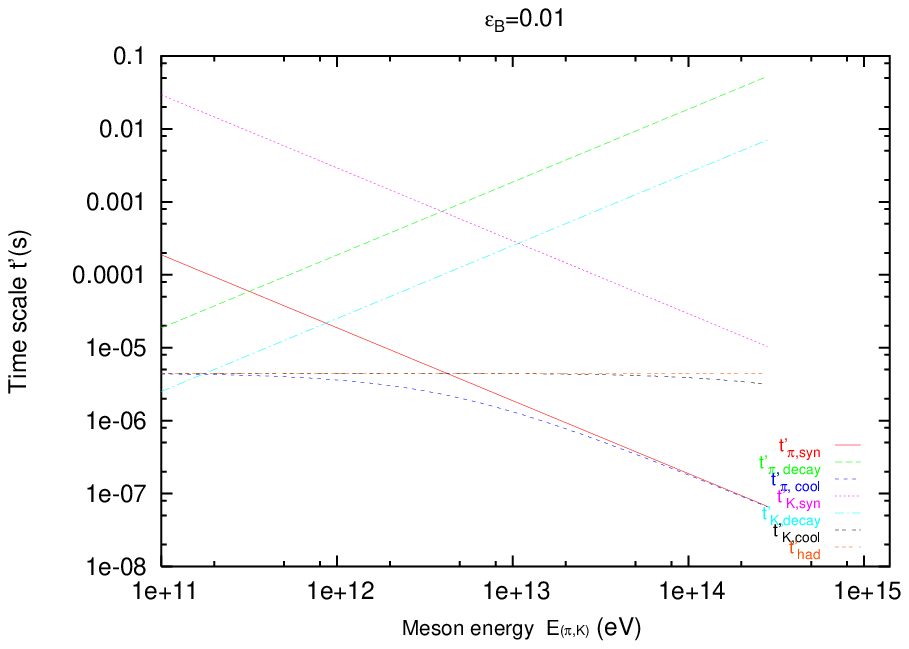}}
\resizebox*{0.45\textwidth}{0.22\textheight}
{\includegraphics{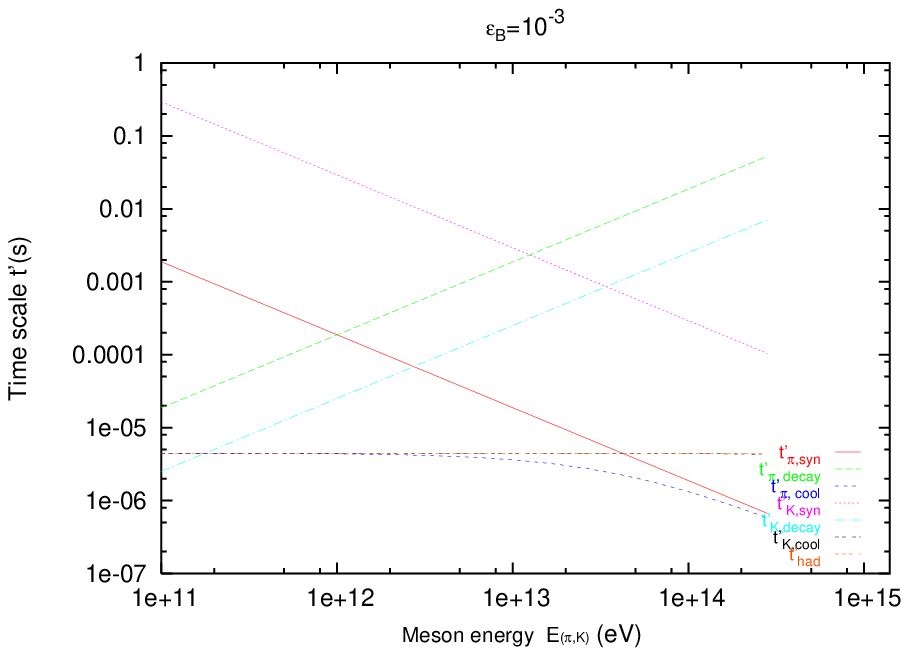}}
\resizebox*{0.45\textwidth}{0.22\textheight}
{\includegraphics{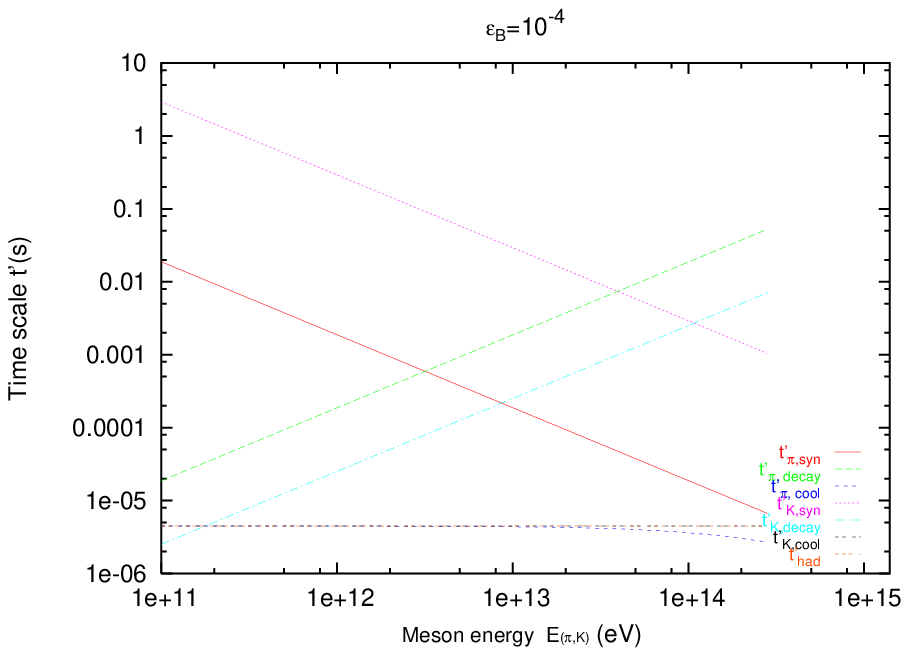}}
}
\caption{Meson $(\pi^{\pm}/K^{\pm})$ cooling time scales in the comoving frame for different processes as a function of Meson energy (E$_{\pi/K}$) when the shell collisions take place  at r=$6\times 10^{10}$ cm and  different magnetic fields.  $\pi/K$- synchrotron radiation (t$'_{(\pi/K),sync}$), $\pi$/K- decay (t$'_{(\pi/K),decay}$),  $\pi$/K- synchroton and hadronic  radiation    (t$'_{cool}$).   }
\label{mtime_r2}
\end{figure}

\begin{figure}
\vspace{0.5cm}
{ \centering
\resizebox*{0.5\textwidth}{0.24\textheight}
{\includegraphics{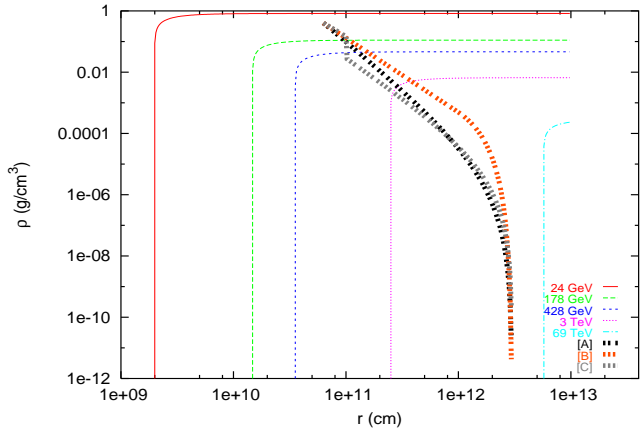}}
 }

\caption{In the top figure, Density profiles  ([A], [B] and [C])  given by eqs. (\ref{dens-pro-A}), (\ref{dens-pro-B}) and (\ref{dens-pro-C}), respectively are plotted.  Also from the resonance condition,  we plot  the resonance density as a function of resonance length for High-energy neutrinos.   In the bottom figure,  the flip probability is plotted as a function of neutrino energy for  density profiles [A] (eq. \ref{dens-pro-A}), [B] (eq. \ref{dens-pro-B}) and [C] (eq. \ref{dens-pro-C}). We have used the best parameters of the three-flavor neutrino oscillation.}
\label{threeflavor}
\end{figure}

\begin{figure}
\vspace{0.5cm}
{ \centering
\resizebox*{0.5\textwidth}{0.29\textheight}
{\includegraphics{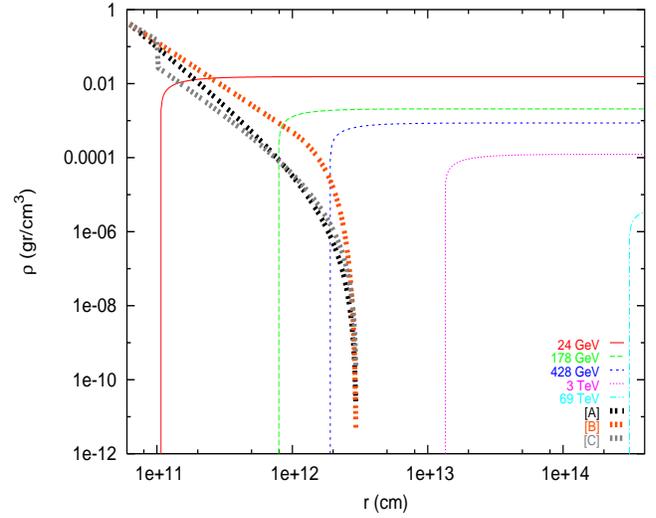}} }
{ \centering
\resizebox*{0.5\textwidth}{0.29\textheight}
{\includegraphics{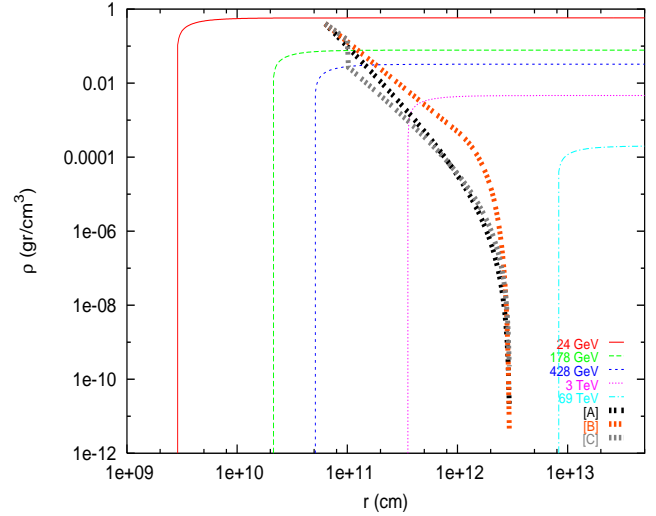}}
}
{ \centering
\resizebox*{0.5\textwidth}{0.29\textheight}
{\includegraphics{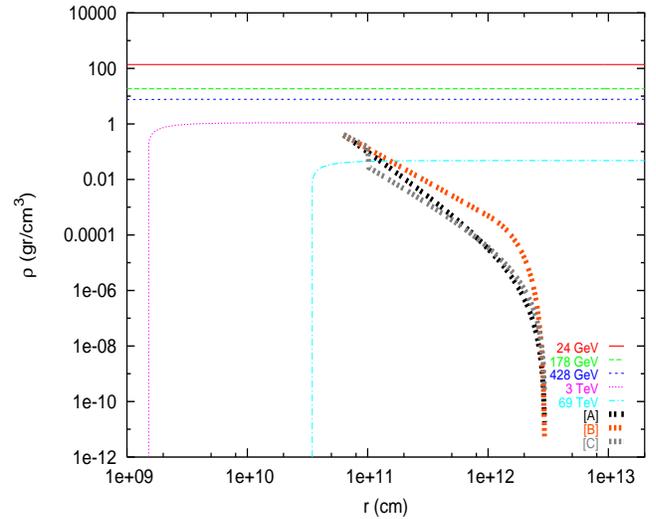}}
}

\caption{Density profiles  ([A], [B] and [C])  given by eqs. (\ref{dens-pro-A}), (\ref{dens-pro-B}) and (\ref{dens-pro-C}), respectively are plotted.  Also from the resonance condition,  we plot  the resonance density as a function of resonance length for High-energy neutrinos. We have used the best parameters of the two-flavor solar (above), atmospheric (medium) and accelerator (below) neutrino oscillation.}
\label{twoflavor}
\end{figure}

\begin{figure}
\vspace{0.5cm}
{ \centering
\resizebox*{0.5\textwidth}{0.25\textheight}
{\includegraphics{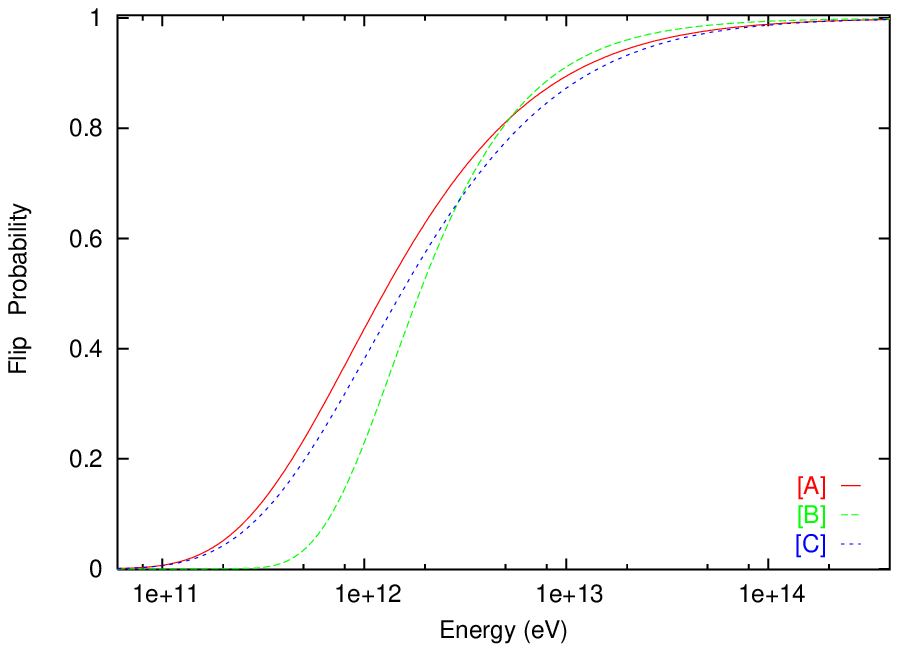}}
\resizebox*{0.5\textwidth}{0.25\textheight}
{\includegraphics{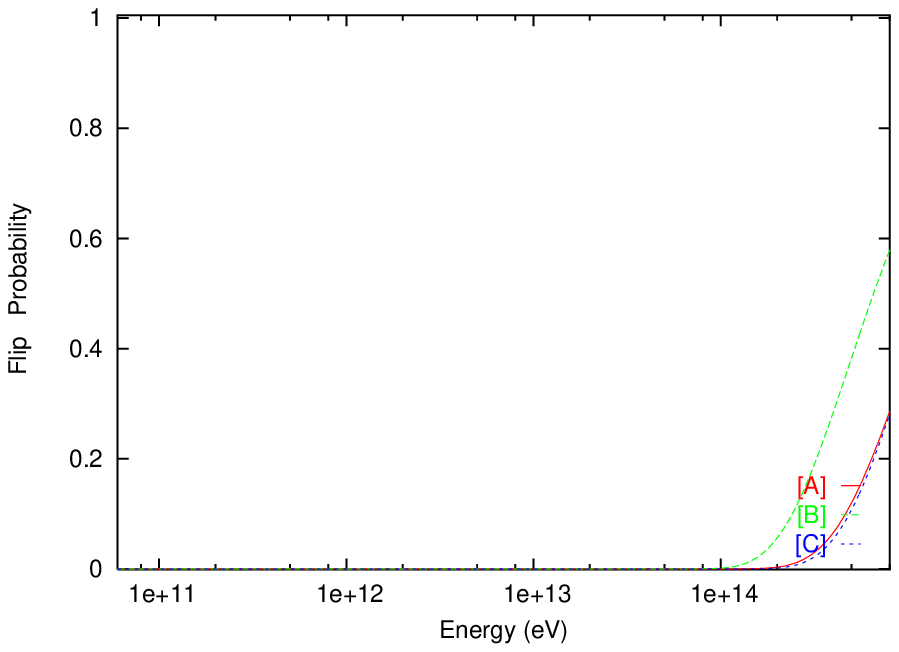}}
\resizebox*{0.5\textwidth}{0.25\textheight}
{\includegraphics{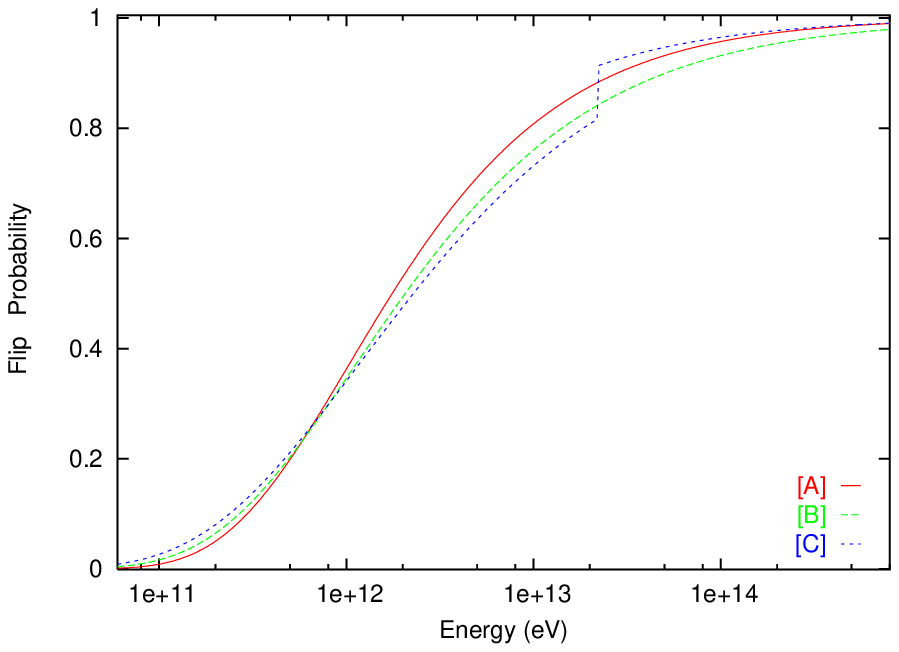}}
}
\caption{The flip probability is plotted as a function of neutrino energy for  density profiles [A] (eq. \ref{dens-pro-A}), [B] (eq. \ref{dens-pro-B}) and [C] (eq. \ref{dens-pro-C}).  On the top figure we use solar parameters, on the middle figure we use atmospheric parameters and on the bottom figure we use accelerator neutrinos. }
\label{twoflip}
\end{figure}

\begin{figure}
\vspace{0.5cm}
{ \centering
\resizebox*{0.5\textwidth}{0.18\textheight}
{\includegraphics{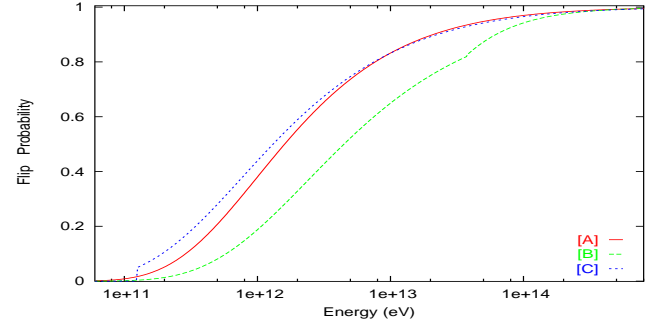}}
 }
\caption{The flip probability is plotted as a function of neutrino energy for  density profiles [A] (eq. \ref{dens-pro-A}), [B] (eq. \ref{dens-pro-B}) and [C] (eq. \ref{dens-pro-C}).  We have used the best parameters of the three-flavor neutrino oscillation.}
\label{threeflip}
\end{figure}

\begin{figure}
\vspace{0.5cm}
{ \centering
\resizebox*{0.5\textwidth}{0.18\textheight}
{\includegraphics{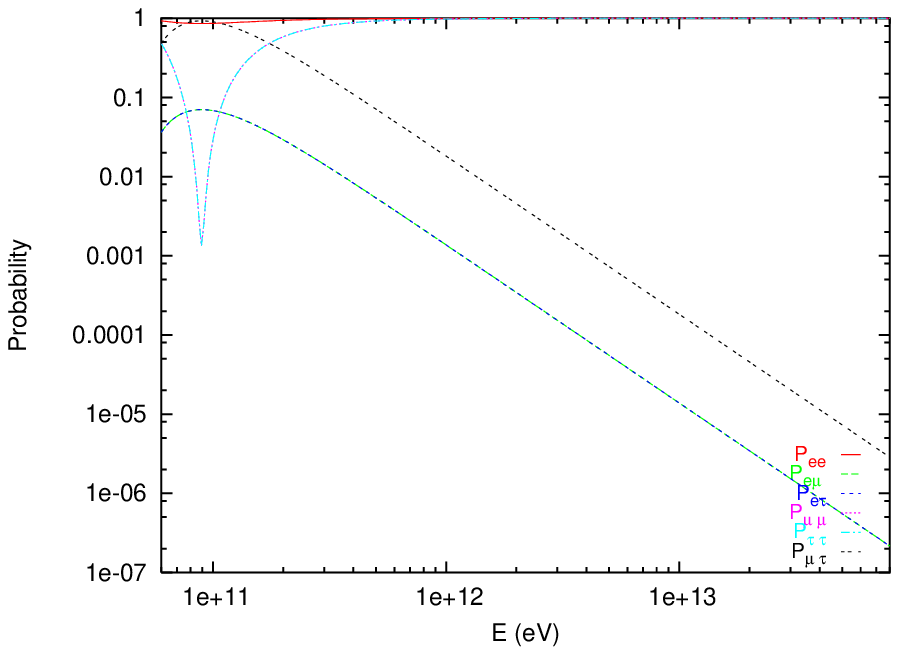}} }
{ \centering
\resizebox*{0.5\textwidth}{0.18\textheight}
{\includegraphics{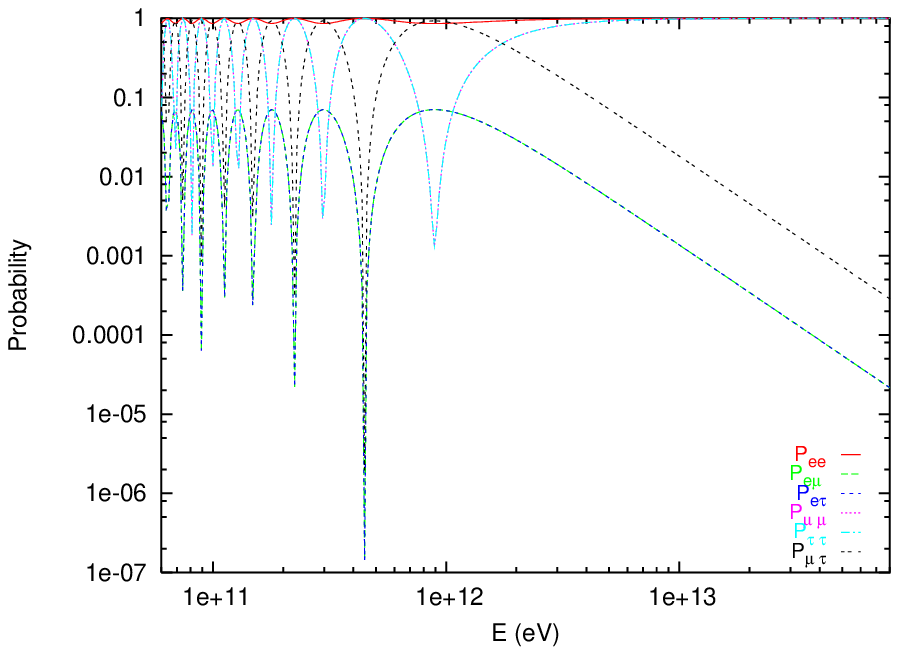}} }
{ \centering
\resizebox*{0.5\textwidth}{0.19\textheight}
{\includegraphics{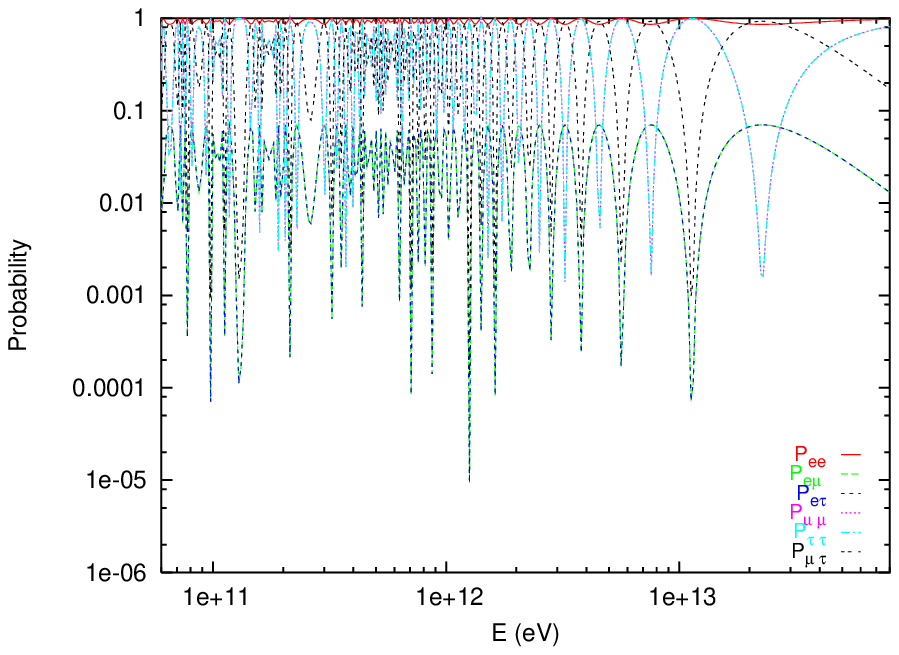}}
}
\caption{We have plotted the oscillation probability as a function of neutrino energy for  neutrinos  produced at radii  $r=6\times10^{9}$ cm and $r=6\times10^{10}$ cm.     In the top, middle and bottom figures  we plot the oscillation probabilities when neutrinos are moving at $r=10^{10}$ cm, $r=10^{11}$ cm  and $r=10^{12}$ cm, respectively.}
\label{proen}
\end{figure}

\begin{figure}
\vspace{0.5cm}
{ \centering
\resizebox*{0.5\textwidth}{0.22\textheight}
{\includegraphics{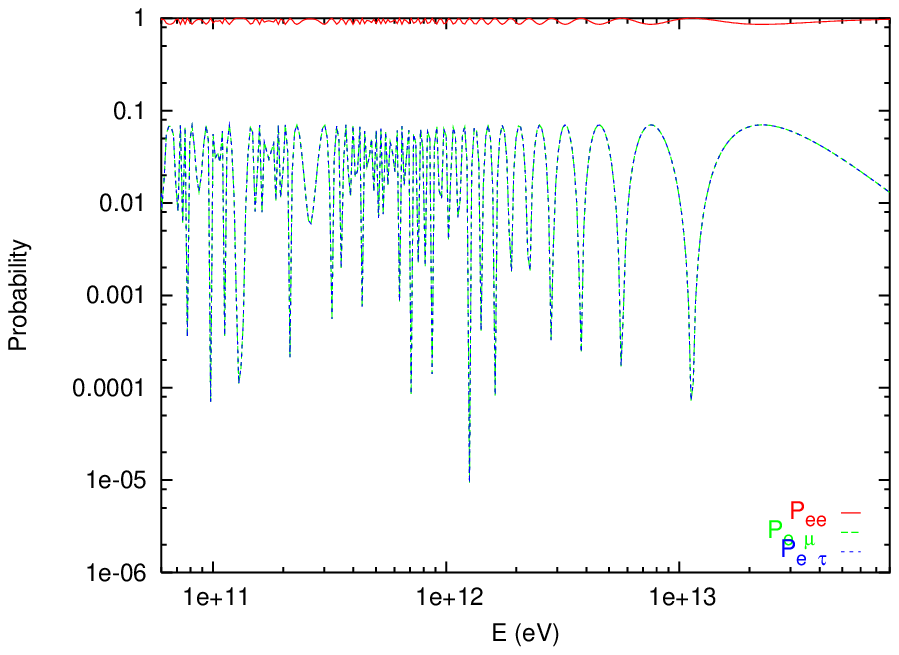}}
\resizebox*{0.5\textwidth}{0.22\textheight}
{\includegraphics{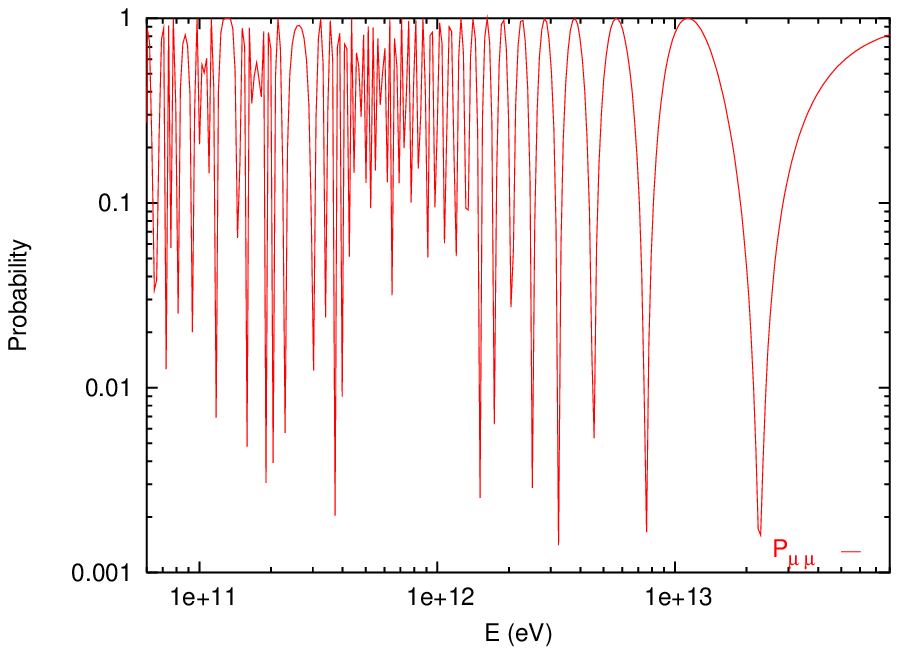}}
\resizebox*{0.5\textwidth}{0.22\textheight}
{\includegraphics{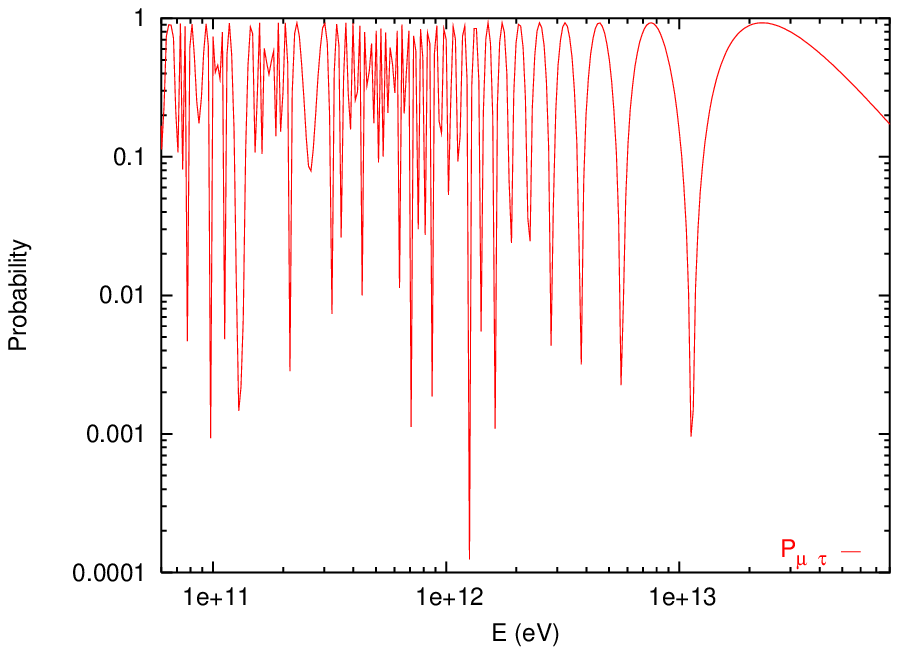}}
\resizebox*{0.5\textwidth}{0.22\textheight}
{\includegraphics{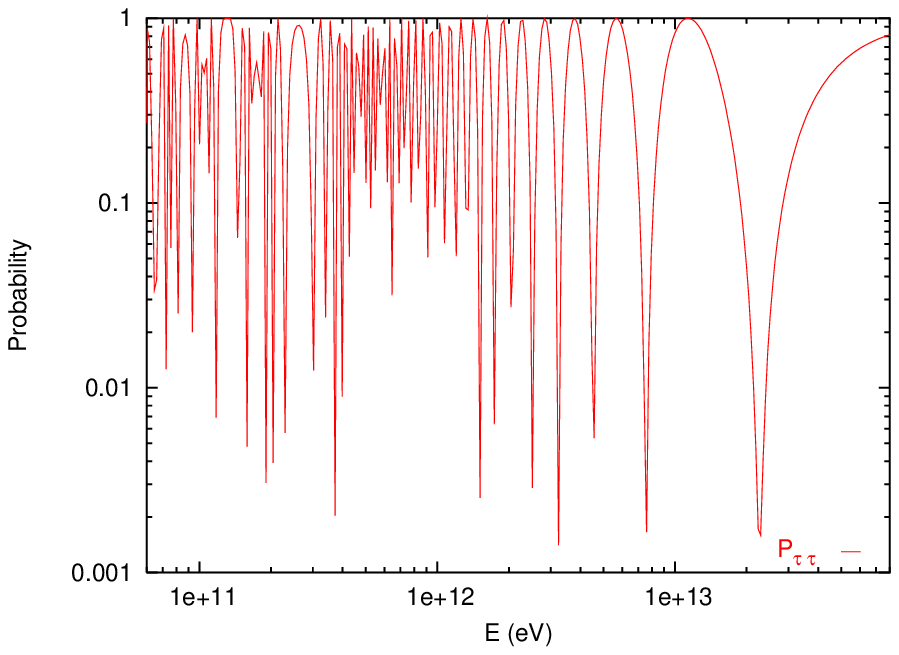}}
}
\caption{To have a better visibility of bottom figure \ref{proen}, we  separate  the oscillation probabilities like, figures from up to down,  $P_{ei}$ i for e, $\mu$, $\tau$ (first one), $P_{\mu\mu}$   (second one),  $P_{\mu\tau}$ (third one)  and  $P_{\tau\tau}$  (four one).  }
\label{prosep}
\end{figure}

\begin{figure}
\vspace{0.5cm}
{ \centering
\resizebox*{0.5\textwidth}{0.22\textheight}
{\includegraphics{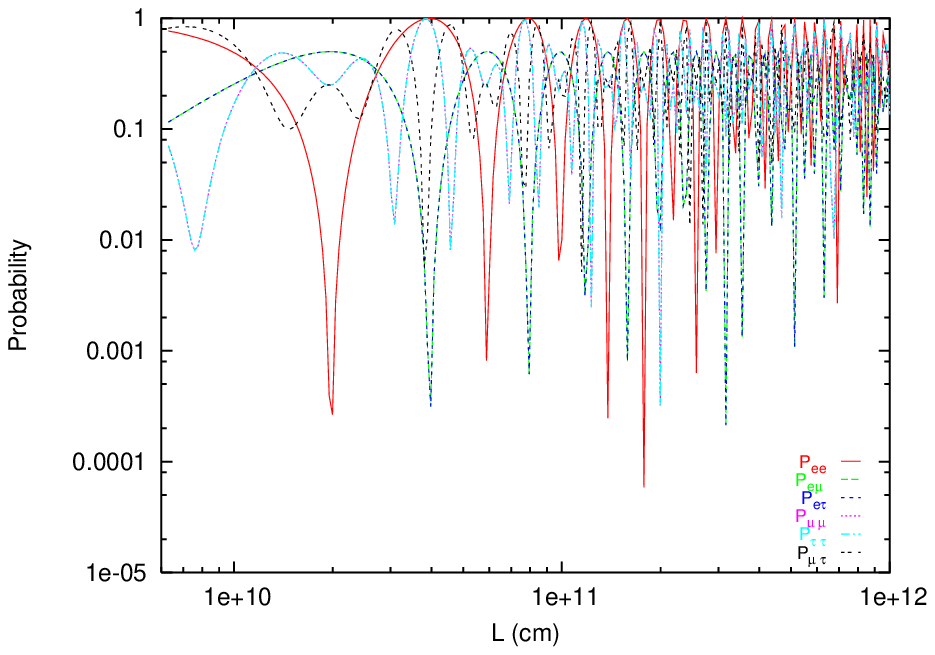}}
\resizebox*{0.5\textwidth}{0.22\textheight}
{\includegraphics{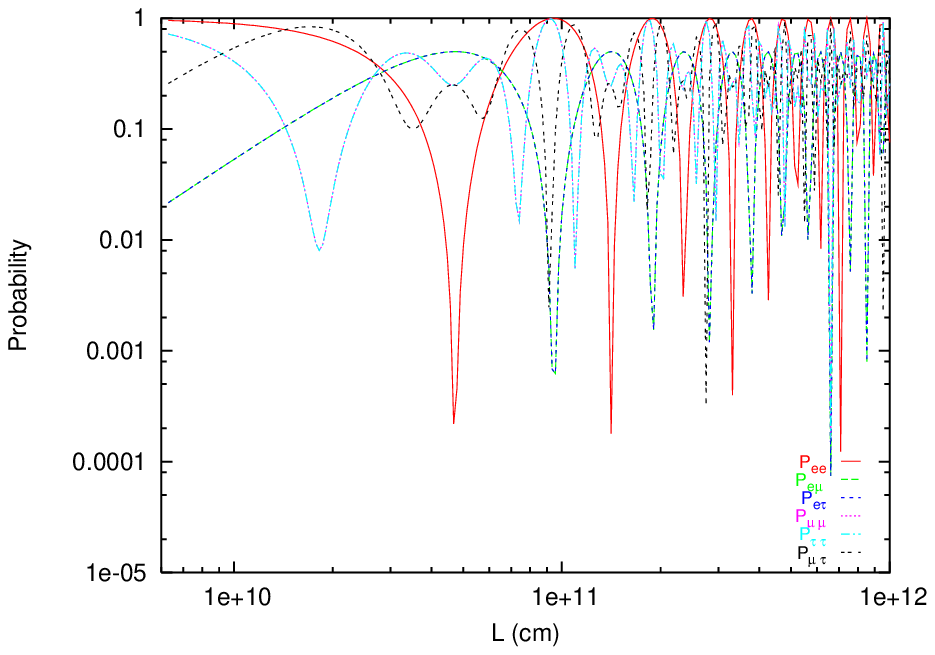}}
\resizebox*{0.5\textwidth}{0.22\textheight}
{\includegraphics{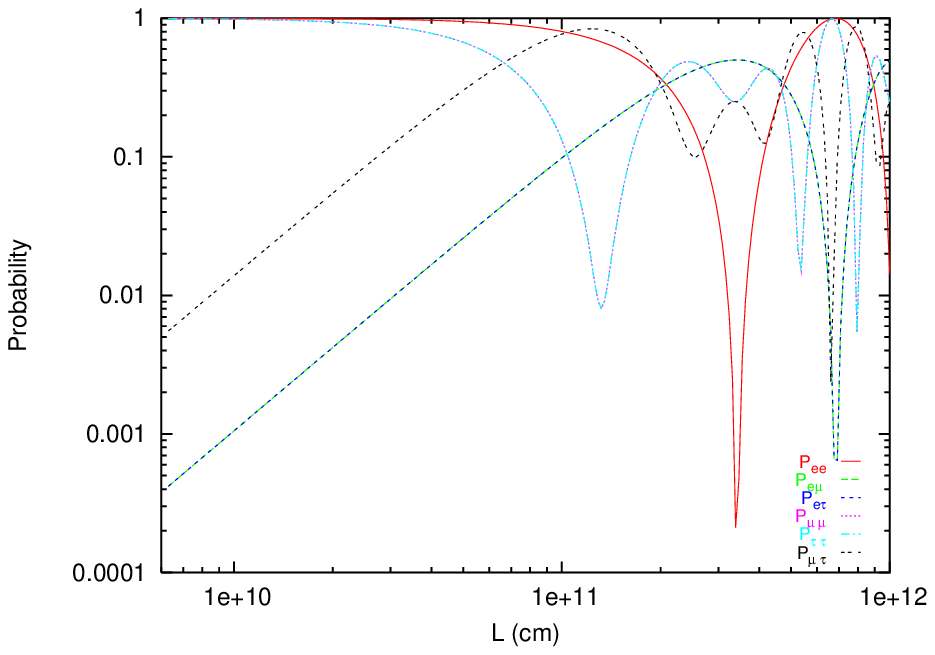}}
\resizebox*{0.55\textwidth}{0.22\textheight}
{\includegraphics{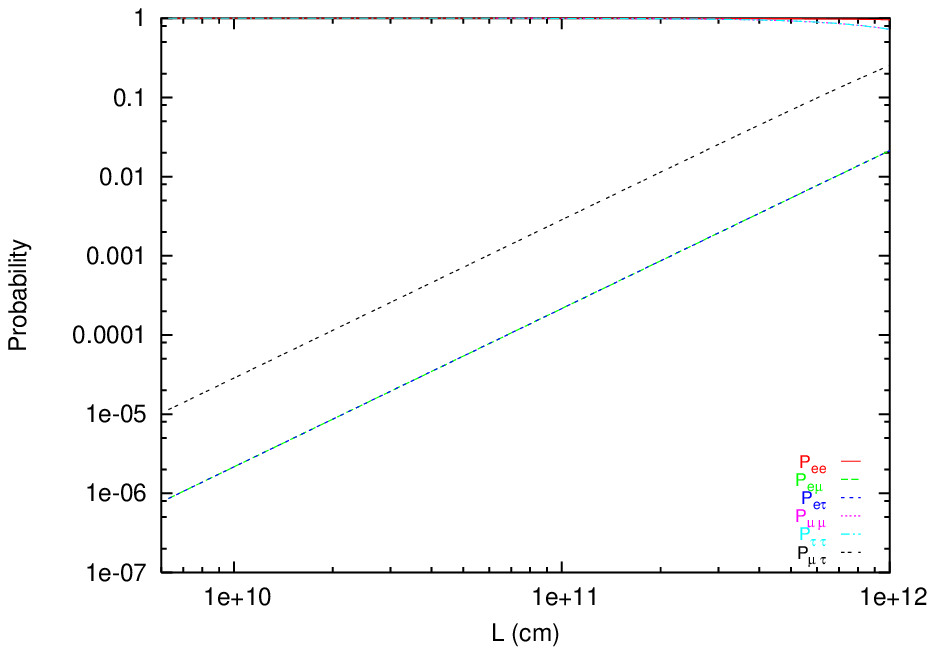}}
}
\caption{We plot the oscillation probability when  neutrinos are  produced at a radius  $r=6\times10^9$ cm and are propagating through  the jet direction for four energies:  $E_\nu=178$ GeV (first figure),  $E_\nu=428$ GeV (second figure), $E_\nu=3$ TeV (third figure) and $E_\nu=69$ TeV (four figure) }
\label{prob_dist}
\end{figure}

\begin{figure}
\vspace{0.5cm}
{ \centering
\resizebox*{0.5\textwidth}{0.22\textheight}
{\includegraphics{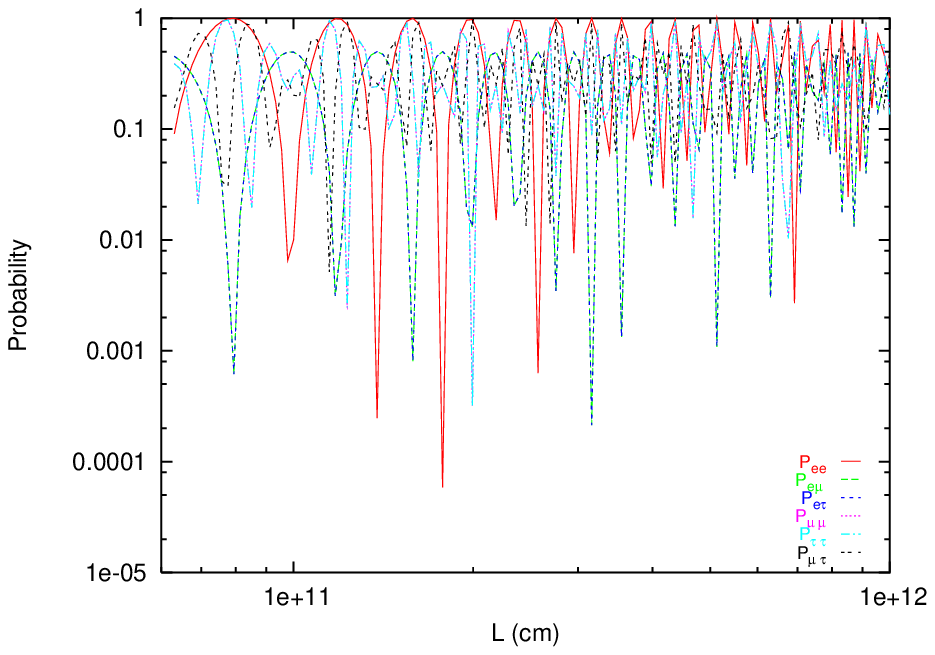}}
\resizebox*{0.5\textwidth}{0.2\textheight}
{\includegraphics{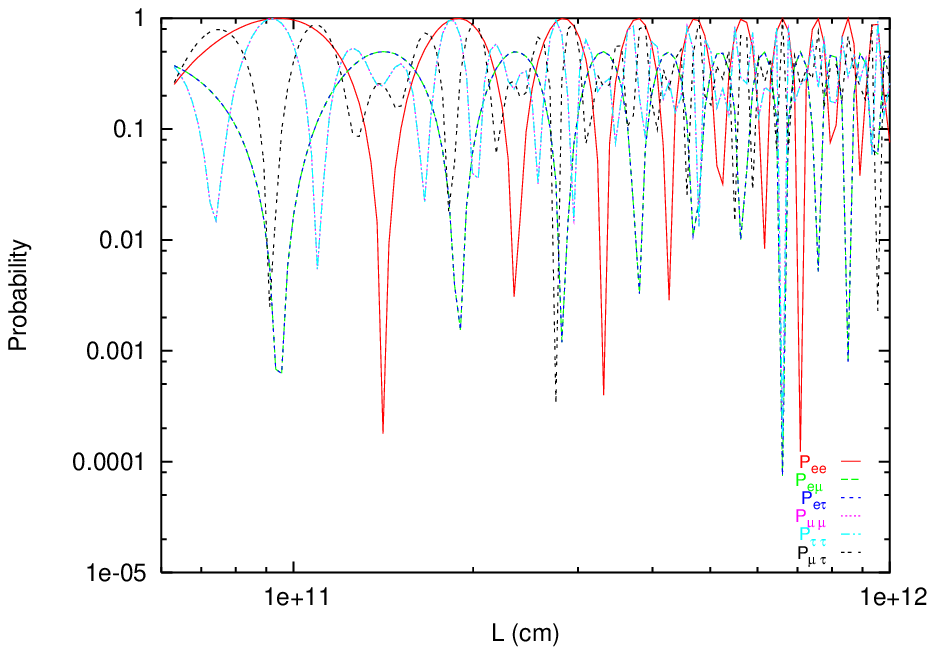}}
\resizebox*{0.5\textwidth}{0.22\textheight}
{\includegraphics{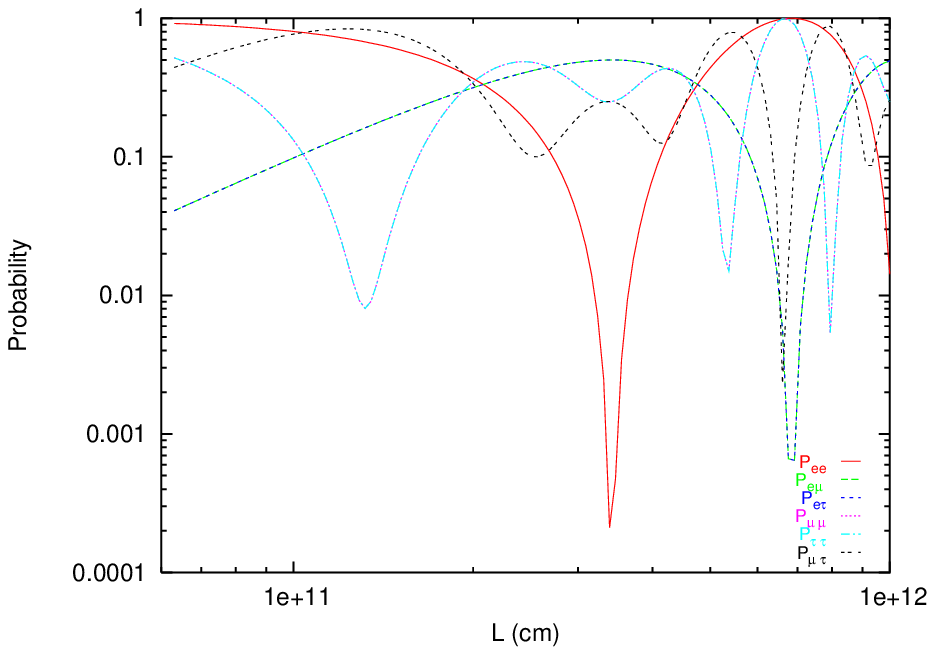}}
\resizebox*{0.5\textwidth}{0.22\textheight}
{\includegraphics{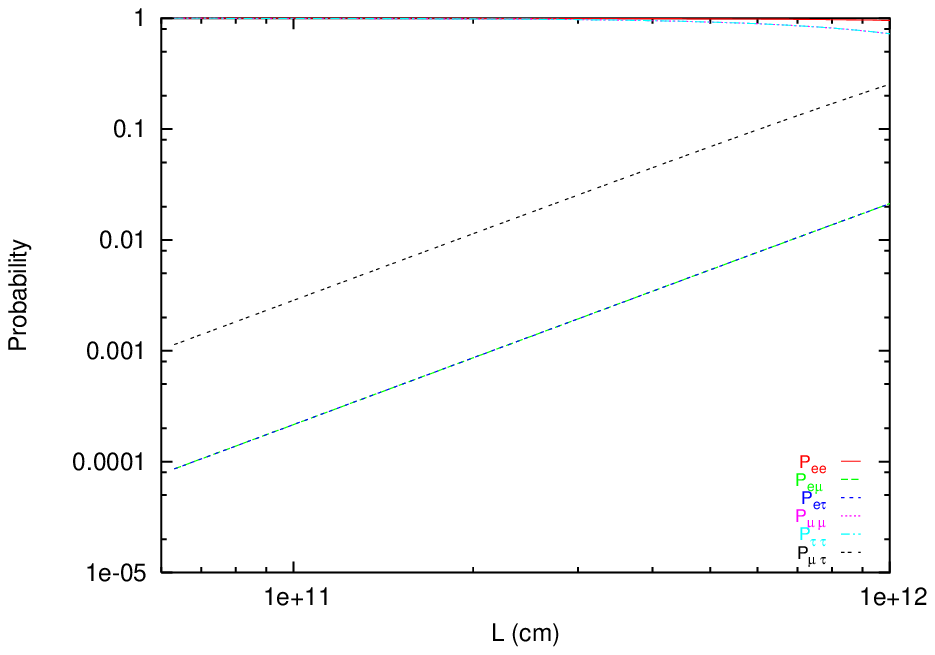}}
}
\caption{We plot the oscillation probability when  neutrinos are  produced at a radius  $r=6\times10^{10}$ cm and are propagating through  the jet direction for four energies:  $E_\nu=178$ GeV (first figure),  $E_\nu=428$ GeV (second figure), $E_\nu=3$ TeV (third figure) and $E_\nu=69$ TeV (four figure) }
\label{prob_dist2}
\end{figure}

We have considered a core collapse of massive stars leading to supernovae (SNe) of type Ib,c and II with mildly relativistic jets.  Although this mildly relativistic jet may not be able to break through the stellar envelope,  electrons and protons are expected to be accelerated in the internal shocks,   and then to be cooled down by synchrotron radiation, inverse Compton and hadronic processes (p$\gamma$ and p-hadron/meson).   Photons from  electron synchrotron radiation thermalized  to a some keV-peak  energy serve  as  cooling mechanism for accelerated protons by means of  p$\gamma$ interactions. Another cooling mechanism of protons considered here are the  p-p interactions, due to  the high number density of protons  (3.1 $\times 10^{20}$ cm$^{-3} \leq n'_p \leq $ 3.1 $\times 10^{22}$ cm$^{-3}$ ) \citep{raz05}.   In p$\gamma$ and p-p interactions, high-energy pions and kaons are created  which  in turn interact with protons by $\pi$-p and $K$-p interactions, producing  another hadronic/meson  cooling mechanism.   To illustrate the degree and  energy region of efficiency of each cooling process, we have plotted the proton (figures \ref{ptime_r1} and \ref{ptime_r2}) and meson (figures \ref{mtime_r1} and \ref{mtime_r2}) time scales when   internal shocks take place at   $6\times 10^{9}$ cm and $6\times 10^{10}$ cm  and, the magnetic field lies in the range  3.4$\times 10^7$ G  $\leq B' \leq$ 1.1$\times 10^{10}$ G.  Comparing  the time scales in figures \ref{ptime_r1} and \ref{ptime_r2}, one can observe that the maximum proton energy is when the acceleration and synchrotron time scales are equal; it happens  when proton energy is in the range $10^{15} eV \leq E'_p \leq 10^{16}$  eV which corresponds to  internal shocks at  $6\times 10^9$ cm  with $B' = 1.1\times 10^{10}$ G   and  $6\times 10^{10}$ cm   with $B'=3.4\times 10^7$ G, respectively.  In figs. \ref{mtime_r1} and \ref{mtime_r2},   one can see that hadronic time scales are equal to  other  time scales at different energies.  For instance,  internal shocks at $6\times 10^{10}$ cm and $B'=1.1\times 10^{9}$ G,  the time scales of pion synchrotron emission and  hadronic are equal  for  pion energy  $\sim 5\times 10^{11}$ eV. Computing the break meson energies for which time scales are equal to each other, we  can estimate the break neutrino energies.   From the equality  of  kaon/pion lifetime and  synchrotron  cooling time scales we obtain the break neutrino energies  $\sim$(24/179) GeV and  $\sim$428 GeV/69 TeV, respectively.  Also, considering   p$\gamma$ interactions  the  threshold neutrino energy $\sim$ 3 TeV is obtained.  Taking into account the distances of  internal shocks  ($6\times 10^{9}$ cm and $6\times 10^{10}$ cm)
we have plotted  the neutrino energy  as a function of the magnetic equipartition parameter in the range  0.1$\leq \epsilon_B\leq 10^{-4}$  (3.4$\times 10^7$ G  $\leq B' \leq$ 1.1$\times 10^{10}$ G).  As shown in the fig.  \ref{prod_neu},  neutrino energy between 1 - 10 PeV can be generated for  $\epsilon_B$  between 3.5$\times 10^{-3}$  and 4.1$\times 10^{-4}$, that corresponds to a magnetic field in the range  2.02$\times 10^8$ (2.02$\times 10^9$) G - 6.9$\times 10^7$ (6.9$\times 10^8$) G at $6\times 10^{9}$ cm and $6\times 10^{10}$ cm from the central engine, respectively. Under this scenario, chocked jets are bright in high-energy neutrinos and dark in gamma rays.\\
On the other hand, taking into account the range of neutrino energy  (24 GeV$\leq E_\nu\leq$ 69 TeV),  internal shocks  at a distance of  6$\times 10^{10}$ cm,  strength of  magnetic field of 1.1$\times 10^{10}$ G   and considering  three models of density profile  (see section III. eqs. \ref{dens-pro-A}, \ref{dens-pro-B} and \ref{dens-pro-C}) of a pre-supernova star,  we  present a full description of two- and three-flavor neutrino oscillations.   Based on these models of density profiles we calculate the effective potential, the resonance condition and,   the resonance length and density.    From the resonance  condition, we obtain the  resonance density ($\rho_R$)  as a function of resonance length ($l_R$) for two (eq. \ref{p1}) and three flavors (eq. \ref{p2}). We overlap the plots of the density profiles as a function of distance with  the resonance conditions (resonance density as a function of resonance length). They are shown  in Fig \ref{twoflavor} (two flavors) and in Fig.  \ref{threeflavor} (three flavors).    For two flavors,   we have taken into account solar (top), atmospheric (middle) and accelerator (bottom) parameters of neutrino experiments.   Using solar parameters, the resonance length is in the range  $\sim (10^{11} -  10^{14.2})$ cm  and resonance density in $\sim  (10^{-2} - 10^{-4})$g/cm$^3$. As can be seen,  neutrinos with energy 24 GeV  are the only ones that meet the resonance condition for all models of density profiles while neutrinos of energy  178 GeV meet marginally the resonance condition just for the model [B].   Neutrinos with other energy cannot meet the resonance condition. Using atmospheric parameters, the resonance length lies in the range $\sim (10^{9.1} -  10^{13.3})$ cm and the resonance density in $\sim  (10^{1} - 10^{-4})$g/cm$^3$.   As shown,  neutrinos in the energy range of 178 GeV - 3 TeV can oscillate many times before leaving the source. Although the resonance length of neutrino with energy 24 GeV is smaller than star radius, the resonance density is greater than other  models.   Using accelerator parameters, the resonance length is less than $\sim 10^{10.2}$ cm and  the resonance density lies in the range  $\sim  (10^{2} - 10^{2})$g/cm$^3$. Although the resonance length is smaller than the star radius for two flavors, the one that meets the resonance density  is the neutrino energy  69 TeV.   For three flavors,  the range of resonance length  is $\sim (10^{9} -  10^{12.5})$ cm  and resonance density is  $\sim  (0.9 - 10^{-4})$g/cm$^3$, presenting a similar behavior to that described by means of atmospheric parameters.   
As the dynamics of resonant transitions  is not only determined by the resonance condition, but also by adiabatic conversion,  we plot the flip probability as a  function of neutrino energy for two  (fig \ref{twoflip}) and three flavors (fig \ref{threeflip}). Dividing the plots of flip probabilities in three regions of less than 0.2,  between 0.2 and 0.8 and greater than 0.8, we have that  in the first case (P$_\gamma \leq$ 0.2), a pure adiabatic conversion  occurs, the last case (P$_\gamma \geq$ 0.8) is a strong violation of adiabaticity  and the intermediate region 0.2 $<$ P$_\gamma$ $<$ 0.8 represents the transition region \citep{dig00}. In Fig. \ref{twoflip},  the top, middle and bottom  plots are obtained  using solar, atmospheric and  accelerator parameters of neutrino oscillations, respectively.    As shown in top figure,  the pure adiabatic conversion occurs when neutrino energy is  less than $5\times 10^{11}$ eV for model [A] and [C] and,   $\sim 10^{12}$ eV for model [B] and, the strong violation of adiabaticity is given for neutrino energy greater than $6\times 10^{12}$ eV  in the three profiles.    In the middle figure, one can see that independently of the profile, neutrinos with energy of less than E$_\nu$=10$^{14}$ eV can have pure adiabatic conversions.    In the bottom figure,  the three models of  density profiles have the same behavior for the whole energy range.   Neutrinos with   energy less than $\sim 10^{11.3}$ eV and  greater than  $\sim 10^{13.2}$ eV present conversion adiabatically pure and strong violation, respectively.   In fig. \ref{threeflip},   the flip probability for three flavors are plotted.   The energy range for each region of P$_\gamma$ changes marginally   according to the model of  density profile.  Neutrinos with E$\sim 10^{12}$ eV are capable of having  pure adiabatic conversion in [B] but not in [A] or [C].  The strong violation of adiabaticity begins when the neutrino energies are E$\sim 10^{13}$ eV and E$\sim 10^{13.8}$ eV, for [A] and [C], respectively. \\
On the other hand,   we have also plotted (fig. \ref{proen}) the oscillation probabilities for three flavors  as a function of energy when neutrinos keep moving at a distance of   r=$10^{11}$ cm (above) and r=$10^{12}$ cm (below) from the core.   In the top figure, the survival probability of electron neutrino, P$_{ee}$, is close to one regardless of  neutrino energy, therefore  the conversion probabilities  P$_{\mu e}$ and P$_{\tau e}$ are close to zero, as shown.  Depending on the neutrino energy, the survival probability of muon and tau neutrino, P$_{\mu \mu}$ and P$_{\tau \tau}$, oscillates between zero and one. For example,   for E$\sim 430$ GeV,  the conversion probability of muon  P$_{\mu \tau}$ is close to zero while the survival probability of muon and tau neutrino, P$_{\mu \mu}$ and P$_{\tau \tau}$,  are close to one, and for E$\sim 1$ TeV  probabilities change dramatically, being P$_{\mu \tau}\sim 1$ and P$_{\tau \tau}$=P$_{\mu \mu}\sim$ 0. In the bottom figure, neutrinos are moving along the jet  at r=10$^{12}$ cm and  although the survival and conversion probabilities  have  similar behaviors to those moving to  r=10$^{11}$ cm,  they  are changing faster.  To have a better understanding, we have separated all probabilities and plotted them in fig.  \ref{prosep}.
From up to down,  the probabilities of  electron neutrino and   survival probability of muon neutrino are shown in the first and second graph, respectively,   and  the conversion and survival probability of tau  neutrino are plotted in the third and four graph, respectively. Moreover, we have plotted in figs. \ref{prob_dist} and \ref{prob_dist2} the oscillation probabilities  as a function of distance, when  neutrinos are produced at a radius  $6\times 10^{9}$ cm and $6\times 10^{10}$ cm, respectively,  and continue to propagate along the jet. We take into account  four neutrino energies $E_\nu$=178 GeV, $E_\nu$=428 GeV, $E_\nu$=3 TeV and $E_\nu$=69 TeV.  
As shown, as neutrino energy  increases, the probabilities oscillate less. For instance,  when an electron neutrino with energy $E_\nu$=178 GeV  propagates along the jet,  the survival probability of electron changes from one at  $\sim 8\times 10^{10}$ cm to zero at $\sim 9.5\times 10^{10}$ cm.   For  $E_\nu$=428 GeV(3 TeV), the survival probabilities  change from one at  $9.1\times 10^{10}$ ($6.0\times 10^{10}$) cm  to zero at $1.8\times 10^{11}$($3.5\times 10^{11}$) cm  and for $E_\nu$=69 TeV, the probability  is constant in this range (greater than $\sim 10^{12}$ cm).  In the last case, neutrino does not oscillate to another flavor during its propagation.
Finally, considering a flux ratio for $\pi$, K and $\mu$ decay  of 1:  2:  0, the density profile [A] and  oscillation probabilities at three distances (10$^{11}$ cm,  10$^{11.5}$ cm and 10$^{12}$ cm), we show in table 1 the flavor ratio on the surface of star. Also, computing the vacuum oscillation effects between the source and Earth (Eq. \ref{matrixosc}),  we estimate and show in table \ref{flaratio}  the flavor ratio expected on Earth when neutrinos emerge  from the star at L=(10$^{11}$,  10$^{11.5}$  and 10$^{12}$) cm .

\begin{table}
\begin{center}\renewcommand{\tabcolsep}{0.2cm}
\renewcommand{\arraystretch}{0.89}
\begin{tabular}{|c|c|c|c|c|c|}\hline
$E_{\nu}$  &$\phi_{\nu_e}:\phi_{\nu_\mu}:\phi_{\nu_\tau}$ &$\phi_{\nu_e}:\phi_{\nu_\mu}:\phi_{\nu_\tau}$&$\phi_{\nu_e}:\phi_{\nu_\mu}:\phi_{\nu_\tau}$  \\
(TeV)&(L=10$^{11}$ cm)&(L=10$^{11.5}$ cm)&(L=10$^{12}$ cm)\\ \hline

0.024     &  0.946:1.949:0.115  &  0.697:1.405:0.899   &  0.881:1.578:0.541  \\\hline

0.178     & 0.510:1.814:0.676  &  0.987:1.386:0.627  &  0.507:1.807:0.686 \\\hline

0.428     & 0.983:1.589:0.428  &  0.659:1.871:0.524   &  0.538:1.721:0.741\\\hline

3            & 0.896:1.212:0.892   &  0.502:1.753:0.744   &  0.501:1.762:0.737 \\\hline

68.5     &  0.999:1.997:0.003   &  0.998:1.972:0.030   &  0.979:1.746:0.275 \\\hline

\end{tabular}
\label{tatm}
\end{center}
\caption{\small\sf The flavor ratio on the surface of source for five neutrino energies  (E$_{\nu}$=24 GeV, 178 GeV, 428 GeV, 3 TeV and 68.5 TeV), leaving the star to three distances L=10$^{11}$ cm, 10$^{11.5}$ cm, and 10$^{12}$ cm.  }
\label{flaratio}
\end{table}

\begin{table}
\begin{center}\renewcommand{\tabcolsep}{0.2cm}
\renewcommand{\arraystretch}{0.89}
\begin{tabular}{|c|c|c|c|c|c|}\hline
$E_{\nu}$  &$\phi^0_{\nu_e}:\phi^0_{\nu_\mu}:\phi^0_{\nu_\tau}$ &$\phi^0_{\nu_e}:\phi^0_{\nu_\mu}:\phi^0_{\nu_\tau}$&$\phi^0_{\nu_e}:\phi^0_{\nu_\mu}:\phi^0_{\nu_\tau}$  \\
(TeV)&(L=10$^{11}$ cm)&(L=10$^{11.5}$ cm)&(L=10$^{12}$ cm)\\ \hline

0.024     &  1.046:1.008:0.956   &   0.925:1.031:1.045 &  0.998:1.011:0.991 \\\hline

0.178     &   0.889:1.049:1.062  &   1.021:1.000:0.978 &  0.888:1.049:1.063 \\\hline

0.428     &   1.033:1.000:0.966  &  0.954:1.053:1.047  &  0.893:1.046:1.061 \\\hline

3             &   0.979:1.010:1.011  &   0.883:1.049:1.067  &  0.883:1.050:1.067 \\\hline

68.5       &   1.065:0.998:0.936  &   1.063:0.999:0.939 &   1.042:1.001:0.957\\\hline

\end{tabular}
\label{tatm}
\end{center}
\caption{\small\sf The flavor ratio expected on Earth for five neutrino energies  (E$_{\nu}$=24 GeV, 178 GeV, 428 GeV, 3 TeV and 68.5 TeV), leaving the star to three distances L=10$^{11}$ cm, 10$^{11.5}$ cm, and 10$^{12}$ cm.}
\label{flaratio}
\end{table}
\section{Summary and conclusions}
We have done a wide description of  production channels  of high-energy neutrinos in a middle relativistic hidden jet and also shown that neutrinos with energies between  1 - 10 PeV can be generated. Taking  into account a particular range of neutrino energies  generated  in the internal shocks at a distance of  6$\times 10^{10}$ cm and with a distribution of  magnetic field  1.1$\times 10^{10}$ G,  we have shown  their oscillations between flavors along the jet for  three models of density profiles.  For two neutrinos mixing,  we have used the fit values of neutrino oscillation parameters from solar, atmospheric, and accelerator experiments and  analyzing the resonance condition we found that  the resonance lengths are the largest  and  resonance densities are the smallest for solar parameters and  using  accelerator parameters we have obtained the opposite situation,  the resonance lengths are the smallest  and  resonance densities are the largest. The most favorable condition for high-energy neutrinos  to oscillate resonantly before going out of the source is given through atmospheric parameters and   these conversions would be pure adiabatic.
For three neutrino mixing,  we have calculated the ratio flavor on the surface of the source as well as that expected on Earth. Our analysis shows that deviations  from 1:1:1 are obtained  at different energies and places along the jet, which is given in table 2.     From analysis of flip probability we also show that neutrinos may oscillate depending on their energy and the parameters of neutrino experiments.    As a particular case, when the three-flavor  parameters are considered (fig. \ref{threeflip}), we obtain that neutrino energies above  $\geq$ 10 TeV can hardly oscillate, obtaining the same result given by \citet{2013arXiv1304.4906O}.\\  As shown,  depending on the flavor ratio obtained on Earth we could differentiate  the progenitor,  its density profile at different depths in the source,  as well as understand similar features between lGRBs and core collapse supernovae.  Distinct times of arrival of neutrino flavor ratio will provide constraints on density profiles at different places in the star  \citep{bar12}. 
These observations in detectors such as IceCube, Antares and KM3Net would be a compelling evidence that chocked jets  are bright in neutrinos \citep{abb12, abb13, pra10,lei12}.
The number of sources with hidden jets  may be much larger than the exhibited one, limited only by the ratio of type Ib/c and type II SNe to GRB rates. Within 10 Mpc, the rate of core-collapse supernovae is $\sim$1 - 3 yr$^{-1}$, with a large contribution of galaxies around  3 - 4 Mpc.   At larger distances,  the expected number of neutrino events in IceCube is still several, and the supernova rate is $\geq$ 10 yr$^{-1}$ at 20 Mpc \citep{and05}. Recently,  \citet{tab10} calculated the events expected in DeepCore and neutrino-induced cascades in km$^3$ detectors for neutrinos energies $\leq$ 10 GeV and $\leq$  a few TeV respectively and forecast that $\sim$ 4 events in DeepCore and $\sim$ 6 neutrino-induced cascades in IceCube/KM3Net  would be expected. An extension up to higher energies of this  calculation  should be done to correlate the expected events  in these sources with  the number of PeV-neutrinos  observed with IceCube \citep{aar13}. \\  
Interference effects in the detector by atmospheric neutrino oscillation are very  small (less than 10 \%) due to short   path traveled by neutrinos in comparison with cosmological distances \citep{men07}.

\section*{Acknowledgements}

We thank the referee  for a critical reading of the paper and valuable suggestions. We also thank B. Zhang, K. Murase, William H. Lee, Fabio de Colle, Enrique Moreno and Antonio Marinelli for useful discussions.   NF gratefully acknowledges a Luc Binette-Fundaci\'on UNAM Posdoctoral Fellowship.


\begin{thebibliography}{}

\bibitem[Abbasi et al. (2007)]{abb12} Abbasi R. et~al., 2012, Astroparticle Physics, 35, 615

\bibitem[{{Abe} \& et~al.(2011)}]{abe11a}
{Abe} K., et~al., 2011, Physical Review Letters, 107, 241801

\bibitem[{{Aguilar-Arevalo} \& et~al.(2007)}]{agu07}
{Aguilar-Arevalo} A.~A., et~al., 2007, Physical Review Letters, 98, 231801

\bibitem[{{Aguilar-Arevalo} \& et~al.(2009)}]{agu09}
{Aguilar-Arevalo} A.~A., et~al., 2009, Physical Review Letters, 102, 101802

\bibitem[{{Aguilar-Arevalo} \& et~al.(2010)}]{agu10}
{Aguilar-Arevalo} A.~A., et~al., 2010, Physical Review Letters, 105, 181801

\bibitem[{{Aharmim} \& et~al.(2011)}]{aha11}
{Aharmim} B., et~al., 2011, ArXiv e-prints

\bibitem[{{Ahn} \& et~al.(2006)}]{ahn06}
{Ahn} M.~H., et~al., 2006, \prd, 74, 072003

\bibitem[{{Akhmedov} {et~al}\mbox{.}(2004){Akhmedov}, {Johansson}, {Lindner},
  {Ohlsson}, \& {Schwetz}}]{akh04}
{Akhmedov} E.~K., {Johansson} R., {Lindner} M., {Ohlsson} T., {Schwetz} T.,
  2004, Journal of High Energy Physics, 4, 78

\bibitem[{{Alvarez-Mu{\~n}iz} \& {Halzen}(2002)}]{alv02}
{Alvarez-Mu{\~n}iz} J., {Halzen} F., 2002, \apjl, 576, L33

\bibitem[{{Alvarez-Mu{\~n}iz} \& {M{\'e}sz{\'a}ros}(2004)}]{alv04}
{Alvarez-Mu{\~n}iz} J., {M{\'e}sz{\'a}ros} P., 2004, \prd, 70, 123001

\bibitem[{{Ando} \& {Beacom}(2005)}]{and05}
{Ando} S., {Beacom} J.~F., 2005, Physical Review Letters, 95, 061103

\bibitem[{{Araki} \& et~al.(2005)}]{ara05}
{Araki} T., et~al., 2005, Physical Review Letters, 94, 081801

\bibitem[{{Armbruster} \& et~al.(2002)}]{arm02}
{Armbruster} B., et~al., 2002, \prd, 65, 112001

\bibitem[{{Arnett}(1991)}]{arn91}
{Arnett} D., 1991, \apj, 383, 295

\bibitem[{{Athanassopoulos} \& et~al.(1996)}]{ath96}
{Athanassopoulos} C., et~al., 1996, Physical Review Letters, 77, 3082

\bibitem[{{Athanassopoulos} \& et~al.(1998)}]{ath98}
{Athanassopoulos} C., et~al., 1998, Physical Review Letters, 81, 1774

\bibitem[{{Athar}, {Je{\.z}abek} \& {Yasuda}(2000){Athar}, {Je{\.z}abek}, \&
  {Yasuda}}]{ath00}
{Athar} H., {Je{\.z}abek} M., {Yasuda} O., 2000, \prd, 62, 103007

\bibitem[Atoyan\& Dermer(2001)]{ato01}Atoyan A., Dermer C. D., 2001, Physical Review Letters, 87, 221102

\bibitem[{{Barniol Duran}, {Bo{\v s}njak} \& {Kumar}(2012){Barniol Duran},
  {Bo{\v s}njak}, \& {Kumar}}]{barn12}
{Barniol Duran} R., {Bo{\v s}njak} {\v Z}., {Kumar} P., 2012, \mnras, 424, 3192

\bibitem[{{Bartos}, {Dasgupta} \& {M{\'a}rka}(2012){Bartos}, {Dasgupta}, \&
  {M{\'a}rka}}]{bar12}
{Bartos} I., {Dasgupta} B., {M{\'a}rka} S., 2012, \prd, 86, 083007

\bibitem[{{Becker} \& {Biermann}(2009)}]{bec09}
{Becker} J.~K., {Biermann} P.~L., 2009, Astroparticle Physics, 31, 138

\bibitem[{{Bethe} \& {Pizzochero}(1990)}]{bet90}
{Bethe} H.~A., {Pizzochero} P., 1990, \apjl, 350, L33

\bibitem[{{Chen}(1985)}]{che85}
{Chen} H.~H., 1985, Physical Review Letters, 55, 1534

\bibitem[{{Chevalier} \& {Soker}(1989)}]{che89}
{Chevalier} R.~A., {Soker} N., 1989, \apj, 341, 867

\bibitem[{{Cholis} \& {Hooper}(2012)}]{cho12}
{Cholis} I., {Hooper} D., 2012, ArXiv e-prints

\bibitem[{{Church} {et~al}\mbox{.}(2002){Church}, {Eitel}, {Mills}, \&
  {Steidl}}]{chu02}
{Church} E.~D., {Eitel} K., {Mills} G.~B., {Steidl} M., 2002, \prd, 66, 013001

\bibitem[{{Costantini} \& {Vissani}(2005)}]{cos05}
{Costantini} M.~L., {Vissani} F., 2005, Astroparticle Physics, 23, 477

\bibitem[{{Dermer} \& {Atoyan}(2003)}]{der03}
{Dermer} C.~D., {Atoyan} A., 2003, Physical Review Letters, 91, 071102

\bibitem[{{Dermer} \& {Menon}(2009)}]{der09}
{Dermer} C.~D., {Menon} G., 2009, {High Energy Radiation from Black Holes:
  Gamma Rays, Cosmic Rays, and Neutrinos}

\bibitem[{{Dighe} \& {Smirnov}(2000)}]{dig00}
{Dighe} A.~S., {Smirnov} A.~Y., 2000, \prd, 62, 033007

\bibitem[{{Fraija}, {Gonz{\'a}lez} \& {Lee}(2012){Fraija}, {Gonz{\'a}lez}, \&
  {Lee}}]{fra12}
{Fraija} N., {Gonz{\'a}lez} M.~M., {Lee} W.~H., 2012, \apj, 751, 33

\bibitem[{{Gonzalez-Garcia}(2011)}]{gon11}
{Gonzalez-Garcia} M.~C., 2011, Physics of Particles and Nuclei, 42, 577

\bibitem[{{Gonzalez-Garcia} \& {Maltoni}(2008)}]{gon08}
{Gonzalez-Garcia} M.~C., {Maltoni} M., 2008, \physrep, 460, 1

\bibitem[{{Gonzalez-Garcia} \& {Nir}(2003)}]{gon03}
{Gonzalez-Garcia} M.~C., {Nir} Y., 2003, Reviews of Modern Physics, 75, 345

\bibitem[{{Gupta} \& {Zhang}(2007)}]{gup07}
{Gupta} N., {Zhang} B., 2007, Astroparticle Physics, 27, 386

\bibitem[{{Hjorth} \& et~al.(2003)}]{hjo03}
{Hjorth} J., et~al., 2003, \nat, 423, 847

\bibitem[{{IceCube Collaboration} {et~al}\mbox{.}(2013{\natexlab{a}}){IceCube
  Collaboration}, {Aartsen}, {Abbasi}, {Abdou}, {Ackermann}, {Adams},
  {Aguilar}, {Ahlers}, {Altmann}, {Auffenberg}, \& et~al.}]{aar13}
{IceCube Collaboration} {et~al.}, 2013{\natexlab{a}}, ArXiv e-prints

\bibitem[{{IceCube Collaboration} {et~al}\mbox{.}(2013{\natexlab{b}}){IceCube
  Collaboration}, {Abbasi}, {Abdou}, {Ackermann}, {Adams}, {Aguilar}, {Ahlers},
  {Altmann}, {Andeen}, {Auffenberg}, \& et~al.}]{abb13}
{IceCube Collaboration} {et~al.}, 2013{\natexlab{b}}, \apj, 763, 33

\bibitem[{{Kashti} \& {Waxman}(2005)}]{kas05}
{Kashti} T., {Waxman} E., 2005, Physical Review Letters, 95, 181101

\bibitem[{{Kumar} \& {Barniol Duran}(2010)}]{kum10}
{Kumar} P., {Barniol Duran} R., 2010, \mnras, 409, 226

\bibitem[{{Kuo} \& {Pantaleone}(1989)}]{kuo89}
{Kuo} T.~K., {Pantaleone} J., 1989, Reviews of Modern Physics, 61, 937

\bibitem[{{Learned} \& {Pakvasa}(1995)}]{lea95}
{Learned} J.~G., {Pakvasa} S., 1995, Astroparticle Physics, 3, 267

\bibitem[{{Leisos} {et~al}\mbox{.}(2012){Leisos}, {Tsirigotis}, {Tzamarias}, \&
  {KM3NeT consortium}}]{lei12}
{Leisos} A., {Tsirigotis} A.~G., {Tzamarias} S.~E., {KM3NeT consortium} f.~t.,
  2012, ArXiv e-prints

\bibitem[{{Liu} \& {Wang}(2013)}]{liu13}
{Liu} R.-Y., {Wang} X.-Y., 2013, \apj, 766, 73

\bibitem[{{MacLachlan} {et~al}\mbox{.}(2013){MacLachlan}, {Shenoy}, {Sonbas},
  {Dhuga}, {Cobb}, {Ukwatta}, {Morris}, {Eskandarian}, {Maximon}, \&
  {Parke}}]{2013MNRAS.432..857M}
{MacLachlan} G.~A. {et~al.}, 2013, \mnras, 432, 857

\bibitem[{{Matzner} \& {McKee}(1999)}]{mat99}
{Matzner} C.~D., {McKee} C.~F., 1999, \apj, 510, 379

\bibitem[{{Mena}, {Mocioiu} \& {Razzaque}(2007){Mena}, {Mocioiu}, \&
  {Razzaque}}]{men07}
{Mena} O., {Mocioiu} I., {Razzaque} S., 2007, \prd, 75, 063003

\bibitem[{{Meszaros}, {Rees} \& {Wijers}(1998){Meszaros}, {Rees}, \&
  {Wijers}}]{mes98}
{Meszaros} P., {Rees} M.~J., {Wijers} R.~A.~M.~J., 1998, \apj, 499, 301

\bibitem[{{Murase} {et~al}\mbox{.}(2006){Murase}, {Ioka}, {Nagataki}, \&
  {Nakamura}}]{mur06}
{Murase} K., {Ioka} K., {Nagataki} S., {Nakamura} T., 2006, \apjl, 651, L5

\bibitem[{{Nellen}, {Mannheim} \& {Biermann}(1993){Nellen}, {Mannheim}, \&
  {Biermann}}]{nell93}
{Nellen} L., {Mannheim} K., {Biermann} P., 1993, \prd, 47, 5270

\bibitem[{{Osorio Oliveros}, {Sahu} \& {Sanabria}(2013){Osorio Oliveros},
  {Sahu}, \& {Sanabria}}]{2013arXiv1304.4906O}
{Osorio Oliveros} A.~F., {Sahu} S., {Sanabria} J.~C., 2013, ArXiv e-prints

\bibitem[{{Pradier} \& {Antares Collaboration}(2010)}]{pra10}
{Pradier} T., {Antares Collaboration}, 2010, Classical and Quantum Gravity, 27,
  194004

\bibitem[{{Razzaque}, {M{\'e}sz{\'a}ros} \& {Waxman}(2004){Razzaque},
  {M{\'e}sz{\'a}ros}, \& {Waxman}}]{raz04b}
{Razzaque} S., {M{\'e}sz{\'a}ros} P., {Waxman} E., 2004, Physical Review
  Letters, 93, 181101

\bibitem[{{Razzaque}, {M{\'e}sz{\'a}ros} \& {Waxman}(2005){Razzaque},
  {M{\'e}sz{\'a}ros}, \& {Waxman}}]{raz05}
{Razzaque} S., {M{\'e}sz{\'a}ros} P., {Waxman} E., 2005, Modern Physics Letters
  A, 20, 2351

\bibitem[{{Razzaque} \& {Smirnov}(2010)}]{raz10}
{Razzaque} S., {Smirnov} A.~Y., 2010, Journal of High Energy Physics, 3, 31

\bibitem[{{Sacahui} {et~al}\mbox{.}(2012){Sacahui}, {Fraija}, {Gonz{\'a}lez},
  \& {Lee}}]{sac12}
{Sacahui} J.~R., {Fraija} N., {Gonz{\'a}lez} M.~M., {Lee} W.~H., 2012, \apj,
  755, 127

\bibitem[{{Sahu} \& {Zhang}(2010)}]{sah10}
{Sahu} S., {Zhang} B., 2010, Research in Astronomy and Astrophysics, 10, 943

\bibitem[{{Shen}, {Kumar} \& {Piran}(2010){Shen}, {Kumar}, \& {Piran}}]{she10}
{Shen} R., {Kumar} P., {Piran} T., 2010, \mnras, 403, 229

\bibitem[{{Shigeyama} \& {Nomoto}(1990)}]{shi90}
{Shigeyama} T., {Nomoto} K., 1990, \apj, 360, 242

\bibitem[{{Shirai} \& {KamLAND Collaboration}(2007)}]{shi07}
{Shirai} J., {KamLAND Collaboration}, 2007, Nuclear Physics B Proceedings
  Supplements, 168, 77

\bibitem[{{Stanek} \& et~al.(2003)}]{sta03}
{Stanek} K.~Z., et~al., 2003, \apjl, 591, L17

\bibitem[{{Stecker}(1968)}]{ste68}
{Stecker} F.~W., 1968, Physical Review Letters, 21, 1016

\bibitem[{{Stecker} {et~al}\mbox{.}(1991){Stecker}, {Done}, {Salamon}, \&
  {Sommers}}]{ste91}
{Stecker} F.~W., {Done} C., {Salamon} M.~H., {Sommers} P., 1991, Physical
  Review Letters, 66, 2697

\bibitem[{{Szabo} \& {Protheroe}(1994)}]{sza94}
{Szabo} A.~P., {Protheroe} R.~J., 1994, Astroparticle Physics, 2, 375

\bibitem[{{Taboada}(2010)}]{tab10}
{Taboada} I., 2010, \prd, 81, 083011

\bibitem[{{the KamLAND Collaboration} \& {Mitsui}(2011)}]{mit11}
{the KamLAND Collaboration}, {Mitsui} T., 2011, Nuclear Physics B Proceedings
  Supplements, 221, 193

\bibitem[{{Wang} {et~al}\mbox{.}(2007){Wang}, {Razzaque}, {M{\'e}sz{\'a}ros},
  \& {Dai}}]{wan07}
{Wang} X.-Y., {Razzaque} S., {M{\'e}sz{\'a}ros} P., {Dai} Z.-G., 2007, \prd,
  76, 083009

\bibitem[{{Waxman} \& {Bahcall}(1997)}]{wax97}
{Waxman} E., {Bahcall} J., 1997, Physical Review Letters, 78, 2292

\bibitem[{{Waxman} \& {Loeb}(2001)}]{wax01}
{Waxman} E., {Loeb} A., 2001, Physical Review Letters, 87, 071101

\bibitem[{{Wendell} \& et~al.(2010)}]{wen10}
{Wendell} R., et~al., 2010, \prd, 81, 092004

\bibitem[{{Wolfenstein}(1978)}]{wol78}
{Wolfenstein} L., 1978, \prd, 17, 2369

\bibitem[{{Woosley}, {Langer} \& {Weaver}(1993){Woosley}, {Langer}, \&  {Weaver}}]{woo93}
{Woosley} S.~E., {Langer} N., {Weaver} T.~A., 1993, \apj, 411, 823



\end{thebibliography}

\clearpage

\end{document}